\documentclass[11pt]{article} 
\usepackage{cite}
\usepackage{amsmath,amsfonts}
\usepackage{bm}
\usepackage{mathrsfs}
\usepackage{latexsym}
\usepackage{amssymb}
\usepackage[breaklinks=true,linktocpage=true]{hyperref}
\hypersetup{colorlinks=true, allcolors=blue}
\usepackage{graphicx}

\usepackage{slashed}
\usepackage{mathbbol}
\usepackage{siunitx}

\usepackage{comment}
\usepackage{physics}
\usepackage{float}

\begin{document}
\pagestyle{empty}
\begin{flushright}
YITP-26-55
\end{flushright}

\vspace*{2.5cm}
\begin{center}

{\bf\LARGE
Impact of matter effects on the unitarity test of lepton mixing
}
\\

\vspace*{1.5cm}
{\large 
Ryuichiro Kitano$^{1}$, 
Joe Sato$^{2}$, and 
Sho Sugama$^{2}$
} \\
\vspace*{0.5cm}

{\it 
$^1$Yukawa Institute for Theoretical Physics, Kyoto University,
Kyoto 606-8502, Japan\\
$^2$Department of Physics, Faculty of Engineering Science, Yokohama National University, Yokohama 240-8501, Japan 
}

\end{center}

\vspace*{1.0cm}

\begin{abstract}
{\normalsize
Testing the unitarity of the lepton mixing matrix, in a manner analogous to the unitarity tests of the CKM matrix in the quark sector, is an important step toward probing physics beyond the standard three-generation framework.
In long baseline neutrino oscillation experiments, the formula of the oscillation probabilities 
can be written as a sum of terms with various combinations of the mixing-matrix elements, and their 
coefficients depend differently on energy.
By observing the spectral information of long baseline experiments such as T2HK and a future neutrino factory at J-PARC with a $\nu_e$ beam, the elements of the mixing matrix can be extracted without assuming a specific parametrization of the mixing matrix.
We investigate how such an extraction method can be applied to neutrino oscillations 
by taking into account matter effects, and discuss how one can test unitarity of the 
mixing matrix in future long baseline experiments.
As a concrete example, we examine the unitarity test by using a four-generation model,
where we look at a quantity which should be vanishing in a unitary model.
Among possible combinations of measurements, the most powerful test can be provided from
the energy spectra of the CP-conjugate appearance channels $\nu_\mu \to \nu_e$ and $\bar{\nu}_\mu \to \bar{\nu}_e$ at T2HK, as well as from the T-conjugate pair $\nu_\mu \to \nu_e$ and $\nu_e \to \nu_\mu$ available at neutrino factories.
}
\end{abstract} 

\newpage
\baselineskip=18pt
\setcounter{page}{2}
\pagestyle{plain}
\baselineskip=18pt
\pagestyle{plain}

\setcounter{footnote}{0}

\section{Introduction}
\label{sec:1}
It has been established that neutrinos undergo oscillation among flavors
at the macroscopic length scale, and various experiments have been measuring the
oscillation probabilities by using solar, atmospheric, reactor, and accelerator neutrinos~\cite{KamLAND:2010fvi,DayaBay:2018yms,SNO:2011hxd,Vinyoles:2016djt,NOvA:2019cyt,OPERA:2018nar,T2K:2018rhz,T2K:2019bcf,KamLAND:2013rgu,Cleveland:1998nv,GNO:2005bds,SAGE:2009eeu,Bellini:2011rx,Super-Kamiokande:2016yck,IceCube:2017lak,IceCube:2019dqi,Super-Kamiokande:2019gzr,JUNO:2015zny}. 
There are many future experiments~\cite{Hyper-Kamiokande:2018ofw,DUNE:2015lol,IceCube-Gen2:2020qha} aiming to further improve the precision of the oscillation parameters.

Most of the analysis has been performed under the assumption of the 
standard three-generation model, and the measurements/discovery 
of the parameters, two mass differences, three mixing angles and a CP phase
are the main targets of the current on-going experiments.
As the next step, it is also important to test fundamental aspects 
such as the CPT theorem and the unitarity of the lepton mixing matrix~\cite{Antusch:2006vwa,Qian:2013ora,Escrihuela:2015wra,Parke:2015goa,Huang:2025znh,Pusty:2026jyw,He:2013rba,Gonzalez-Garcia:2014bfa,He:2016dco,Esteban:2016qun,Ellis:2020ehi,Ellis:2020hus,Guigue:2026dqb}
by over-constraining the parameter space
just as in the quark sector of the Standard Model.
The unitarity of the CKM matrix~\cite{Cabibbo:1963yz,Kobayashi:1973fv} is 
most often represented by the closure of the triangles in the complex plane~\cite{UTfit:2006vpt,Hocker:2001xe,HFLAV:2016hnz,Wolfenstein:1983yz,ParticleDataGroup:2018ovx,Charles:2004jd,Buras:1994ec}.
Similar analysis for the PNMS matrix~\cite{Pontecorvo:1967fh,Maki:1962mu}
should be able to be performed by using the precision measurements in future experiments~\cite{Hyper-Kamiokande:2018ofw,DUNE:2015lol,IceCube-Gen2:2020qha,Antusch:2006vwa,Qian:2013ora,Escrihuela:2015wra,Parke:2015goa,Huang:2025znh,Pusty:2026jyw,He:2013rba,Gonzalez-Garcia:2014bfa,He:2016dco,Esteban:2016qun,Ellis:2020ehi,Ellis:2020hus,Guigue:2026dqb}.

In general, the $3 \times 3$ part of the lepton mixing matrix can be non-unitary especially if there are more than three generations of neutrinos~\cite{Wyler:1982dd,Langacker:1988up,Hewett:1988xc,Buchmuller:1991tu,Ingelman:1993ve,Nardi:1993ag,Chang:1994hz,Tommasini:1995ii,Loinaz:2003gc}.
Also, the check of unitarity would be a quite fundamental test
of quantum field theory. Whether the neutrino masses are obeyed by
the standard rule of the quantum field theory such as unitarity and CPT theorem is an
important question to ask.
If there is some violation of unitarity, the standard parametrization 
by three mixing angle and a phase 
would fail to describe the various oscillation probabilities.
Even without combining different oscillation probabilities, the energy dependence
of a particular mode of oscillation probabilities can differ
from the standard case.
Indeed, testing the unitarity by using the energy dependence 
in future long baseline
neutrino oscillation
experiments
has been proposed in Refs.~\cite{Sato:2000wv,Kitano:2025wpc}, where one extracts 
combinations of the elements of the lepton mixing matrix
from the energy dependence of the neutrino oscillation probability and constructs
a quantity which should 
vanish when the unitarity holds.

For the next generation of long baseline experiments like T2HK, the influence of matter effects becomes non-negligible due to the high precision required for testing unitarity violation. Neglecting these effects, even on relatively short baselines, could potentially lead to a `false positive' signal that erroneously suggests a violation of unitarity. It is therefore crucial to develop a method that accounts for matter effects while maintaining a parametrization-independent approach.

In this paper, we construct a method for testing the unitarity of the lepton mixing matrix as in Ref.~\cite{Kitano:2025wpc} using the energy dependence of the neutrino oscillations in matter.
The presence of the matter effects make the energy dependence quite complicated. Thus, a different analysis from that in vacuum is required.
We assume the T2HK~\cite{Hyper-Kamiokande:2018ofw} experiment and a neutrino factory with a $\nu_e$ beam at J-PARC~\cite{Hamada:2022mua,Kitano:2024kdv,Kitano:2025wpc}, and perform combined analyses of the beams from these experiments.

This paper is organized as follows. In Sec.~\ref{sec:2}, we review neutrino oscillations in matter and present the relation between the neutrino oscillations and the unitarity of the lepton mixing matrix.
In Sec.~\ref{sec:3}, we perform a statistical analysis to test the unitarity of the matrix.
Finally, in Sec.~\ref{sec:4}, we summarize our results.

\section{Neutrino oscillation and Unitarity of the lepton mixing matrix}
\label{sec:2}
The lepton mixing matrix describes the transition between the flavor and the mass eigenstates. 
In the standard Majorana or Dirac three-generation scenarios, 
the lepton mixing matrix (the PMNS matrix) is a $3\times3$ unitary matrix.
In a more general extension of the Standard Model, 
the lepton mixing matrix can be a $(3+M)\times (3+N)$ matrix and its $3 \times 3$ submatrix is
not necessarily a unitary matrix.

In the standard $3\times3$ unitary case, i.e., $U^\dag U = U U^\dag = \vb{1}$,
there are nine unitarity conditions among which six of them correspond to triangles in the complex plane.
These triangles are generally referred to as unitarity triangles, and the unitarity can be tested by checking whether the triangle closes in the complex plane.

In this section, we extend the formulation of Refs.~\cite{Sato:2000wv,Kitano:2025wpc}, 
which discusses only the case
with oscillation in vacuum, to the one with the matter effects. 
We introduce new quantities $\tilde{\xi}$ and $\tilde{\eta}$ which 
can be extracted from the energy dependence of the neutrino oscillation probabilities. 
Since they should vanish 
for $3 \times 3$ unitary PMNS matrix, one can use them as non-trivial test of unitarity.

\subsection{Neutrino oscillation}
In the standard three flavor neutrino oscillations, the time evolution for neutrinos are described by the following equation,
\begin{align}
    i\dv{}{t}\mqty(\nu_e(\bar{\nu}_e) \\ \nu_\mu(\bar{\nu}_\mu) \\ \nu_\tau(\bar{\nu}_\tau))=\tilde{\mathcal{H}}^{(\pm)}
    \mqty(\nu_e(\bar{\nu}_e) \\ \nu_\mu(\bar{\nu}_\mu) \\ \nu_\tau(\bar{\nu}_\tau))~.
    \label{eq:time_evolutiton}
\end{align}
The Hamiltonian for the neutrino flavor evolution is denoted by 
$\tilde{\mathcal{H}}^{(+)}$, while that for antineutrinos is denoted by 
$\tilde{\mathcal{H}}^{(-)}$. The Hamiltonian in matter is given by 
\begin{align}
    \tilde{\mathcal{H}}^{(\pm)} \equiv U^{(*)}{\rm diag}\qty(0,~{\it \Delta}E_{21},~{\it \Delta}E_{31}) U^{\dag({\sf T})}
    \pm {\rm diag}\qty(A,~0,~0)~,
    \label{eq:effective_hamiltonian}
\end{align}
where ${\it \Delta}E_{jk}\equiv {\it \Delta}m_{jk}^2/2E$ denotes the difference of the energy eigenvalues in vacuum. $A=\sqrt{2}G_F n_e$ represents the matter potential. For simplicity, the electron density 
$n_e$ is taken to be constant in our analysis.
Therefore, the neutrino oscillation probability for a baseline $L$ is given by
\begin{align}
    P^{\nu_{\alpha \to \beta}}
    \qty(P^{\bar{\nu}_{\alpha \to \beta}})
    =& \delta_{\alpha \beta} - 4\sum_{j>k}{\rm Re}\qty[\tilde{U}^{(\pm)}_{\alpha j} \tilde{U}^{(\pm)*}_{\beta j}
    \tilde{U}^{(\pm)*}_{\alpha k} \tilde{U}^{(\pm)}_{\beta k}]
    \sin^2\qty(\frac{{\it \Delta}\tilde{E}^{(\pm)}_{jk}L}{2})
    \notag \\
    & - 2\sum_{j>k}{\rm Im}\qty[\tilde{U}^{(\pm)}_{\alpha j} \tilde{U}^{(\pm)*}_{\beta j}
    \tilde{U}^{(\pm)*}_{\alpha k} \tilde{U}^{(\pm)}_{\beta k}]
    \sin\qty({\it \Delta}\tilde{E}^{(\pm)}_{jk}L)
    \notag \\
    =& \delta_{\alpha \beta} 
    - 4{\rm Re}\qty[\tilde{X}^{\alpha \beta(\pm)}_{3} \tilde{X}^{\alpha \beta(\pm)*}_{2}]
    \sin^2\qty(\tilde{\Delta}^{(\pm)}_{32})
    \notag \\
    &- 4{\rm Re}\qty[\tilde{X}^{\alpha \beta(\pm)}_{3} \tilde{X}^{\alpha \beta(\pm)*}_{1}]
    \sin^2\qty(\tilde{\Delta}^{(\pm)}_{31})
    \notag \\
    &- 4{\rm Re}\qty[\tilde{X}^{\alpha \beta(\pm)}_{2} \tilde{X}^{\alpha \beta(\pm)*}_{1}]
    \sin^2\qty(\tilde{\Delta}^{(\pm)}_{21})
    \notag \\
    & - 8 \tilde{J}^{(\pm)}
    \sin\qty(\tilde{\Delta}^{(\pm)}_{32})
    \sin\qty(\tilde{\Delta}^{(\pm)}_{31})
    \sin\qty(\tilde{\Delta}^{(\pm)}_{21})
    \notag \\
    =& \delta_{\alpha \beta} 
    - 4\left| \tilde{X}_3^{\alpha \beta(\pm)} \right|^2
    \sin^2\qty(\tilde{\Delta}^{(\pm)}_{31})
    \notag \\
    &+ 4{\rm Re}\qty[\tilde{X}^{\alpha \beta(\pm)}_{3} \tilde{X}^{\alpha \beta(\pm)*}_{2}]
    \left\{ \sin^2\qty(\tilde{\Delta}^{(\pm)}_{31})-\sin^2\qty(\tilde{\Delta}^{(\pm)}_{32})\right\}
    \notag \\
    &- 4{\rm Re}\qty[\tilde{X}^{\alpha \beta(\pm)}_{2} \tilde{X}^{\alpha \beta(\pm)*}_{1}]
    \sin^2\qty(\tilde{\Delta}^{(\pm)}_{21})
    \notag \\
    & - 8 \tilde{J}^{(\pm)}
    \sin\qty(\tilde{\Delta}^{(\pm)}_{32})
    \sin\qty(\tilde{\Delta}^{(\pm)}_{31})
    \sin\qty(\tilde{\Delta}^{(\pm)}_{21})
    \label{eq:osc_prob}
\end{align}
where $\tilde{U}^{(\pm)}$ is the matrix that diagonalizes $\tilde{\mathcal{H}}^{(\pm)}$, $\tilde{\Delta}_{jk}^{(\pm)} \equiv {\it \Delta}\tilde{E}_{jk}^{(\pm)} L/2$, with ${\it \Delta}\tilde{E}_{jk}^{(\pm)} \equiv \tilde{E}_{j}^{(\pm)} - \tilde{E}_{k}^{(\pm)}$, and $\tilde{E}_{j}^{(\pm)}$ are the eigenvalues of $\tilde{\mathcal{H}}^{(\pm)}$.
The eigenvalues can be derived analytically since the characteristic equation is cubic, and the diagonalization matrix can be obtained in the form 
$\tilde{X}_j^{\alpha \beta (\pm)} \equiv \tilde{U}_{\alpha j}^{(\pm)} \tilde{U}_{\beta j}^{(\pm)*}$ using the method of Kimura, Takamura, and Yokomakura~\cite{Kimura:2002hb,Kimura:2002wd,Yasuda:2007jp}.
$\tilde{J}^{(\pm)}$ is the modified Jarkskog factor in matter~\cite{Harrison:1999df,Naumov:1991ju,Naumov:1991rh}. This is related to the Jarlskog invariant in vacuum, $J=(c_{13}/8)\sin2\theta_{12}\sin2\theta_{13}\sin2\theta_{23}\sin\delta$~\cite{Jarlskog:1985cw,Jarlskog:1985ht}, as follows.
\begin{align}
    \tilde{J}^{(\pm)} &=
    \frac{{\it \Delta}E_{21}{\it \Delta}E_{31}{\it \Delta}E_{32}}
    {{\it \Delta}\tilde{E}_{21}^{(\pm)}{\it \Delta}\tilde{E}_{31}^{(\pm)}{\it \Delta}\tilde{E}_{32}^{(\pm)}}J
    \notag \\
    &=
    \frac{{\it \Delta}E_{21}{\it \Delta}E_{31}{\it \Delta}E_{32}}
    {{\it \Delta}\tilde{E}_{21}^{(\pm)}{\it \Delta}\tilde{E}_{31}^{(\pm)}{\it \Delta}\tilde{E}_{32}^{(\pm)}}
    {\rm Im}\qty[X_3^{\mu e}X_2^{\mu e*}]~,
    \label{eq:jarlskog}
\end{align}
where $X_j^{\alpha \beta} \equiv U_{\alpha j}U_{\beta j}^*$.

Specifically, for the oscillation $\nu_{\mu \to e}$, the expressions are given as follows. 
\begin{align}
	\left| \tilde{X}_3^{\mu e(\pm)} \right|^2 =&
	\frac{1}
	{\qty(\mathit{\Delta}\tilde{E}_{31}^{(\pm)})^2\qty(\mathit{\Delta}\tilde{E}_{32}^{(\pm)})^2}
    \notag \\
    &\times
	\left[ \left|X_3^{\mu e}\right|^2 \mathit{\Delta}E_{31}^2
	\qty(\tilde{E}_1^{(\pm)}+\tilde{E}_2^{(\pm)}-\mathit{\Delta}E_{31}\mp A)^2
	\right.
	\notag \\
	&+\left|X_2^{\mu e}\right|^2 \mathit{\Delta}E_{21}^2
	\qty(\tilde{E}_1^{(\pm)}+\tilde{E}_2^{(\pm)}-\mathit{\Delta}E_{21}\mp A)^2
	\notag \\
	&+2\mathrm{Re}\qty[X_3^{\mu e}X_2^{\mu e*}]\mathit{\Delta}E_{21}\mathit{\Delta}E_{31}
    \notag \\
	&\times \left. \qty(\tilde{E}_1^{(\pm)}+\tilde{E}_2^{(\pm)}-\mathit{\Delta}E_{21}\mp A)
	\qty(\tilde{E}_1^{(\pm)}+\tilde{E}_2^{(\pm)}-\mathit{\Delta}E_{31}\mp A)\right]~,
    \label{eq:C1_mat}
\end{align}
%
\begin{align}
	\mathrm{Re}\qty[\tilde{X}_3^{\mu e(\pm)}\tilde{X}_2^{\mu e(\pm)*}]=&
	\frac{-1}
	{\mathit{\Delta}\tilde{E}_{21}^{(\pm)}\mathit{\Delta}\tilde{E}_{31}^{(\pm)}\qty(\mathit{\Delta}\tilde{E}_{32}^{(\pm)})^2}
    \notag \\
    &\times
	\left[ 
		\left|X_3^{\mu e}\right|^2 \mathit{\Delta}E_{31}^2
	\qty(\tilde{E}_1^{(\pm)}+\tilde{E}_2^{(\pm)}-\mathit{\Delta}E_{31}\mp A)
	\qty(\tilde{E}_1^{(\pm)}+\tilde{E}_3^{(\pm)}-\mathit{\Delta}E_{31} \mp A)
	\right.
	\notag \\
	&+\left|X_2^{\mu e}\right|^2 \mathit{\Delta}E_{21}^2
	\qty(\tilde{E}_1^{(\pm)}+\tilde{E}_2^{(\pm)}-\mathit{\Delta}E_{21}\mp A)
	\qty(\tilde{E}_1^{(\pm)}+\tilde{E}_3^{(\pm)}-\mathit{\Delta}E_{21}\mp A)
	\notag \\
	&+2\mathrm{Re}\qty[X_3^{\mu e}X_2^{\mu e*}]
	\mathit{\Delta}E_{21}\mathit{\Delta}E_{31}
	\left\{\qty(\tilde{E}_1^{(\pm)}+\tilde{E}_2^{(\pm)})\qty(\tilde{E}_1^{(\pm)}+\tilde{E}_3^{(\pm)}) \right.
    \notag \\
	& \left. +\qty(\mathit{\Delta}E_{21}\pm A)\qty(\mathit{\Delta}E_{31}\pm A)\right\}
	\notag \\
	&\left. -\mathrm{Re}\qty[X_3^{\mu e}X_2^{\mu e*}]
		\mathit{\Delta}E_{21}\mathit{\Delta}E_{31}
		\qty(2\tilde{E}_1^{(\pm)}+\tilde{E}_2^{(\pm)}+\tilde{E}_3^{(\pm)})
		\qty(\mathit{\Delta}E_{21}+\mathit{\Delta}E_{31}\pm A)
	\right]~,
    \label{eq:C2_mat}
\end{align}
%
\begin{align}
	\mathrm{Re}\qty[\tilde{X}_2^{\mu e(\pm)}\tilde{X}_1^{\mu e(\pm)*}]=&
	\frac{-1}
	{\qty(\mathit{\Delta}\tilde{E}_{21}^{(\pm)})^2\mathit{\Delta}\tilde{E}_{31}^{(\pm)}\mathit{\Delta}\tilde{E}_{32}^{(\pm)}}
    \notag \\
    &\times
	\left[ 
		\left|X_3^{\mu e}\right|^2 \mathit{\Delta}E_{31}^2
	\qty(\tilde{E}_1^{(\pm)}+\tilde{E}_3^{(\pm)}-\mathit{\Delta}E_{31}\mp A)
	\qty(\tilde{E}_2^{(\pm)}+\tilde{E}_3^{(\pm)}-\mathit{\Delta}E_{31}\mp A)
	\right.
	\notag \\
	&+\left|X_2^{\mu e}\right|^2 \mathit{\Delta}E_{21}^2
	\qty(\tilde{E}_1^{(\pm)}+\tilde{E}_3^{(\pm)}-\mathit{\Delta}E_{21}\mp A)
	\qty(\tilde{E}_2^{(\pm)}+\tilde{E}_3^{(\pm)}-\mathit{\Delta}E_{21}\mp A)
	\notag \\
	&+2\mathrm{Re}\qty[X_3^{\mu e}X_2^{\mu e*}]
	\mathit{\Delta}E_{21}\mathit{\Delta}E_{31}
	\left\{\qty(\tilde{E}_1^{(\pm)}+\tilde{E}_3^{(\pm)})\qty(\tilde{E}_2^{(\pm)}+\tilde{E}_3^{(\pm)}) \right.
    \notag \\
	& \left. +\qty(\mathit{\Delta}E_{21}\pm A)\qty(\mathit{\Delta}E_{31}\pm A)\right\}
	\notag \\
	&\left. -\mathrm{Re}\qty[X_3^{\mu e}X_2^{\mu e*}]
		\mathit{\Delta}E_{21}\mathit{\Delta}E_{31}
		\qty(\tilde{E}_1^{(\pm)}+\tilde{E}_2^{(\pm)}+2\tilde{E}_3^{(\pm)})
		\qty(\mathit{\Delta}E_{21}+\mathit{\Delta}E_{31}\pm A)
	\right]~.
    \label{eq:C3_mat}
\end{align}
Since this paper considers only the oscillations between $\nu_\mu$ and $\nu_e$ (and their anti-neutrinos), the expressions given above are sufficient.
From Eqs.~\eqref{eq:jarlskog}, \eqref{eq:C1_mat}, \eqref{eq:C2_mat}, and \eqref{eq:C3_mat}, the standard neutrino oscillation probability is expressed as a linear combination of four coefficients and four energy-dependent functions,
\begin{align}
    P^{\nu_{\mu \to e}}\qty(P^{\bar{\nu}_{\mu \to e}}) =& 4\left| X_3^{\mu e} \right|^2 
    \cdot \tilde{B}_1^{(\pm)}(E,~X_1^{ee},~X_2^{ee},~X_3^{ee})
    \notag \\
    &+ 4{\rm Re}\qty[X_3^{\mu e} X_2^{\mu e*}]
    \cdot \tilde{B}_2^{(\pm)}(E,~X_1^{ee},~X_2^{ee},~X_3^{ee})
    \notag \\
    &+ 4\left| X_2^{\mu e} \right|^2 
    \cdot \tilde{B}_3^{(\pm)}(E,~X_1^{ee},~X_2^{ee},~X_3^{ee})
    \notag \\
    & + 8{\rm Im}\qty[X_3^{\mu e} X_2^{\mu e*}]
    \cdot \tilde{B}_4^{(\pm)}(E,~X_1^{ee},~X_2^{ee},~X_3^{ee})~.
    \label{eq:osc_prob_linear_combination}
\end{align}
Since the energy-dependent functions are complicated, we denote them by $\tilde{B}_i^{(\pm)}(E)~(i=1\sim4)$.
Note that these four energy-dependent functions also depend on the PMNS matrix elements $X_1^{ee},~X_2^{ee},~X_3^{ee}$ .

\subsection{Implications of neutrino oscillations for the unitarity of the lepton mixing matrix}
The fact that the oscillation probability can be expressed as a linear combination of four coefficients and four energy-dependent functions originates from the assumption that the lepton mixing matrix is the PMNS matrix, i.e., a $3\times 3$ unitary matrix. 
This means that, in general, the neutrino oscillation data cannot be fit by
this function if the actual theory behind is not the standard $3 \times 3$ scenario.
Nevertheless, we proceed with this formula, and derive below a quantity which indicates the unitarity violation
within the $3 \times 3$ scheme if 
the quantity is non vanishing.

From Eq.~\eqref{eq:osc_prob_linear_combination}, seven independent parameters $C_i~(i=1\sim7)$ 
can be extracted by using the energy dependence of the oscillation probability:
\begin{align}
    P^{\nu_{\mu \to e}}(P^{\bar{\nu}_{\mu \to e}})
    =& C_1 \cdot \tilde{B}^{(\pm)}_1(E,~C_5,~C_6,~C_7)
    \notag \\
    &+C_2 \cdot \tilde{B}^{(\pm)}_2(E,~C_5,~C_6,~C_7)
    \notag \\
    &+C_3 \cdot \tilde{B}^{(\pm)}_3(E,~C_5,~C_6,~C_7)
    \notag \\
    &+C_4 \cdot \tilde{B}^{(\pm)}_4(E,~C_5,~C_6,~C_7)~.
    \label{eq:osc_prob_general}
\end{align}
Here we introduce a quantity $\tilde{\xi}$ and $\tilde{\eta}$ as follows
\begin{align}
    \tilde{\xi} \equiv C_1 C_3 - C_2^2 -\frac{C_4^2}{4}~,
    \label{eq:xi}
    \\
    \tilde{\eta} \equiv 1-C_5-C_6-C_7~.
    \label{eq:eta}
\end{align}
For the $3\times3$ unitary PMNS matrix, one finds
\begin{align}
    C_1 &= 4\left| X_3^{\mu e} \right|^2~,
    \label{eq:C1}
    \\
    C_2 &= 4{\rm Re}\qty[X_3^{\mu e} X_2^{\mu e*}]~,
    \label{eq:C2}
    \\
    C_3 &= 4\left| X_2^{\mu e} \right|^2 ~,
    \label{eq:C3}
    \\
    C_4 &= 8{\rm Im}\qty[X_3^{\mu e} X_2^{\mu e*}]~,
    \label{eq:C4}
    \\
    C_5 &= X_1^{ee}~,
    \label{eq:C5}
    \\
    C_6 &= X_2^{ee}~,
    \label{eq:C6}
    \\
    C_7 &= X_3^{ee}~,
    \label{eq:C7}
\end{align}
with which both $\tilde{\xi}$ and $\tilde{\eta}$ vanish.
Note that the correspondence of $C_3$ in Eq.~\eqref{eq:C3} differs from that in Ref.~\cite{Kitano:2025wpc}. For this reason, we use $\tilde{\xi}$ with a slight modification from Ref.~\cite{Kitano:2025wpc}.
The non-vanishing values of $\tilde \xi$ and $\tilde \eta$ immediately indicates the unitarity violation.
Indeed, the test of the $\tilde{\xi}$ parameter corresponds to that of using the unitarity triangle. 
It is worth noting that while $\tilde{\xi}$ and $\tilde{\eta}$ provide distinct criteria for unitarity, $\tilde{\xi}$ is particularly sensitive to the closure of the unitarity triangle in the complex plane. In the following sections, we focus on the extraction of $\tilde{\xi}$ to demonstrate the impact of matter effects on the sensitivity of these tests in future long baseline setups.

\section{Statistical analysis}
\label{sec:3}
We here discuss the methods of extracting the quantities $\tilde{\xi},~\tilde{\eta}$ and evaluate how well the unitarity violation can be tested.
As a simple example, we consider a $4\times4$ unitary model with an eV-scale sterile neutrino.
In the presence of the matter effects, $\tilde{\eta}$ can in principle be obtained experimentally. However, in this paper we do not attempt to extract $\tilde{\eta}$. The justification for this will be discussed later.

\subsection{The least squares method}
    As shown in Eq.~\eqref{eq:osc_prob_general} in Sec.~\ref{sec:2}, the oscillation probability in matter $P(E)$ at a given energy $E$ can be expressed as a linear combination of $C_{1-4}$ and $\tilde{B}^{(\pm)}_{1-4} (E,~C_{5-7})$. Note that the parameters $C_{5-7}$ are non-linear.

    In general, the oscillation probability in a given discrete energy bin $j~(j=1\sim n)$ is expressed as a linear combination of $l$ parameters and energy-dependent functions, with $m$ non-linear parameters.
    \begin{align}
        P(E_j) = \sum_{k=1}^{l} C_k \tilde{B}_k (E_j,~\vb*{C}^{\rm NL})~.
        \label{eq:osc_prob_general_energy_bin}
    \end{align}
    Here we denote the $m$ non-linear parameters as $\vb*{C}^{\rm NL}$. We express Eq.~\eqref{eq:osc_prob_general_energy_bin} as a form of vector and matrix as:
    \begin{align}
        \vb*{P}= \tilde{\mathscr{B}}(\vb*{C}^{\rm NL}) \vb*{C}^{\rm L}~,
        \label{eq:osc_prob_vector}
    \end{align}
    where 
    \begin{align}
        \vb*{P}\equiv \mqty(P(E_1)\\ \vdots \\ P(E_n))
        ,~ \vb*{C}^{\rm L} \equiv \mqty(C_1 \\ \vdots \\ C_l)
        ,~ \tilde{\vb*{B}}_k \equiv \mqty(\tilde{B}_k(E_1,~\vb*{C}^{\rm NL})\\ \vdots \\ \tilde{B}_k(E_n,~\vb*{C}^{\rm NL}))~,
        \label{eq:def1}
    \end{align}
    and
    \begin{align}
        \tilde{\mathscr{B}}(\vb*{C}^{\rm NL}) \equiv \mqty(\tilde{\vb*{B}}_1 & \cdots & \tilde{\vb*{B}}_l)~.
        \label{eq:def2}
    \end{align}
    Here, we explicitly indicate the dependence on the non-linear parameters. Then, similar to Ref.~\cite{Kitano:2025wpc}, the $\chi^2$ value can be defined by using Eq.~\eqref{eq:osc_prob_vector} and  the observed oscillation probability $P^{\rm obs}_j$ at a given energy bin, as follows.
    \begin{align}
        \chi^2\qty(\vb*{C}^{\rm L},~\vb*{C}^{\rm NL}) = \qty(\vb*{P}^{\rm obs}-\mathscr{\tilde{B}}(\vb*{C}^{\rm NL})\vb*{C}^{\rm L})^{\sf T}W
        \qty(\vb*{P}^{\rm obs}-\mathscr{\tilde{B}}(\vb*{C}^{\rm NL})\vb*{C}^{\rm L})~,
        \label{eq:def_chi2}
    \end{align}
    where
    \begin{align}
        \vb*{P}^{\rm obs} \equiv \mqty(P_1^{\rm obs} \\ \vdots \\ P_n^{\rm obs}),~
        W \equiv {\rm diag}\qty(\frac{1}{\qty({\it \Delta}P_1^{\rm obs})^2},~\cdots,~\frac{1}{\qty({\it \Delta}P_n^{\rm obs})^2})~.
        \label{eq:def3}
    \end{align}
    The ${\it \Delta}P_j^{\rm obs}$ is the statistical error of the observed oscillation probability in each energy bin.
    The linear parameter $\vb*{C}^{\rm L}$ to minimize the $\chi^2$ for any $\vb*{C}^{\rm NL}$, namely best fit point of $\vb*{C}^{\rm L}$, is given by
    \begin{align}
        \vb*{C}^{\rm L}_{\rm b.f.p.} \qty(\vb*{C}^{\rm NL})
        = \qty(\tilde{\mathscr{B}}^{\sf T}(\vb*{C}^{\rm NL})W\tilde{\mathscr{B}}(\vb*{C}^{\rm NL}))^{-1}\tilde{\mathscr{B}}^{\sf T}(\vb*{C}^{\rm NL})W
        \vb*{P}^{\rm obs}
        \label{eq:linear_fit}
    \end{align}
    Using this expression, the $\chi^2$ can be written in terms of only $\vb*{C}^{\rm NL}$ as follows,
    \begin{align}\chi^2 \qty(\vb*{C}^{\rm NL}) &=
        \chi^2 \qty(\vb*{C}^{\rm L} =\vb*{C}^{\rm L}_{\rm b.f.p.}\qty(\vb*{C}^{\rm NL}),~\vb*{C}^{\rm NL}) 
        \notag \\
        &= \qty(\vb*{P}^{\rm obs})^{\sf T}
        \qty(\vb{1}_n-\Pi^{\sf T}\qty(\vb*{C}^{\rm NL}))W\qty(\vb{1}_n-\Pi\qty(\vb*{C}^{\rm NL}))\vb*{P}^{\rm obs}~,
        \label{eq:chi2_non_linear}
    \end{align}
    where $\Pi\qty(\vb*{C}^{\rm NL}) \equiv \tilde{\mathscr{B}}\qty(\tilde{\mathscr{B}}^{\sf T}(\vb*{C}^{\rm NL})W\tilde{\mathscr{B}}(\vb*{C}^{\rm NL}))^{-1}\tilde{\mathscr{B}}^{\sf T}(\vb*{C}^{\rm NL})W$.
    By using this expression, one can determine
    $\vb*{C}^{\rm NL}$ that minimizes $\chi^2$. Then, we can obtain $\vb*{C}^{\rm L}$ using the non-linear values. This method is known as variable projection~\cite{doi:10.1137/0710036}.

    When multiple channels from different beams are combined, the parameters can be obtained as in Ref.~\cite{Kitano:2025wpc}.
    However, although it is not mentioned in the reference, there is an important feature to be noted when using CPT-conjugate combinations in the vacuum case.
    For example, when we consider the combination $\qty(\bar{\nu}_{\mu\to e})+\qty(\nu_{e\to\mu})$ in vacuum, both oscillation probabilities have exactly the same energy dependence.
    If the same energy bins are used in the analysis, the energy-dependent functions for each energy bin take the following values, as shown in Eq.~\eqref{eq:def1},
    \begin{align}
        \tilde{\mathscr{B}} = \mqty(\tilde{B}^{(-)}_1(E_1,~-A) & \cdots & \tilde{B}^{(-)}_4(E_1,~-A)
        \\
        & \vdots &
        \\
        \tilde{B}^{(-)}_1(E_n,~-A) & \cdots & \tilde{B}^{(-)}_4(E_n,~-A)
        \\
        \tilde{B}^{(+)}_1(E_1,~+A) & \cdots & \tilde{B}^{(+)}_4(E_1,~+A)
        \\
        & \vdots &
        \\
        \tilde{B}^{(+)}_1(E_n,~+A) & \cdots & \tilde{B}^{(+)}_4(E_n,~+A))~,
        \label{eq:basis_matrix_vac}
    \end{align}
    where $\tilde{B}^{(\pm)}_k(E_j,~\pm A)$ is the energy-dependent function at a given energy bin $E_j$ for neutrino~($+$) and anti-neutrino~($-$).
    Here, the notation is used to make the presence of matter effects explicit.
    Since we now consider the vacuum case, the matter effects are absent, i.e. $A=0$. Consequently, $\tilde{B}^{(+)}_k(E_j)=\tilde{B}^{(-)}_k(E_j)$, and the matrix in Eq.~\eqref{eq:basis_matrix_vac} becomes rank deficient.
    Therefore, the analytical solution in Eq.~\eqref{eq:linear_fit} is not uniquely determined, and the uncertainty is intrinsically larger than in the CP- or T-conjugate combinations.
    Additionally, a similar rank deficiency also arises when $\delta_{\rm CP}=\ang{0}$ or $\ang{180}$.
    
\subsection{Number of events at Hyper-Kamiokande}
\begin{figure}[H]
    \centering
    \includegraphics[width=15cm]{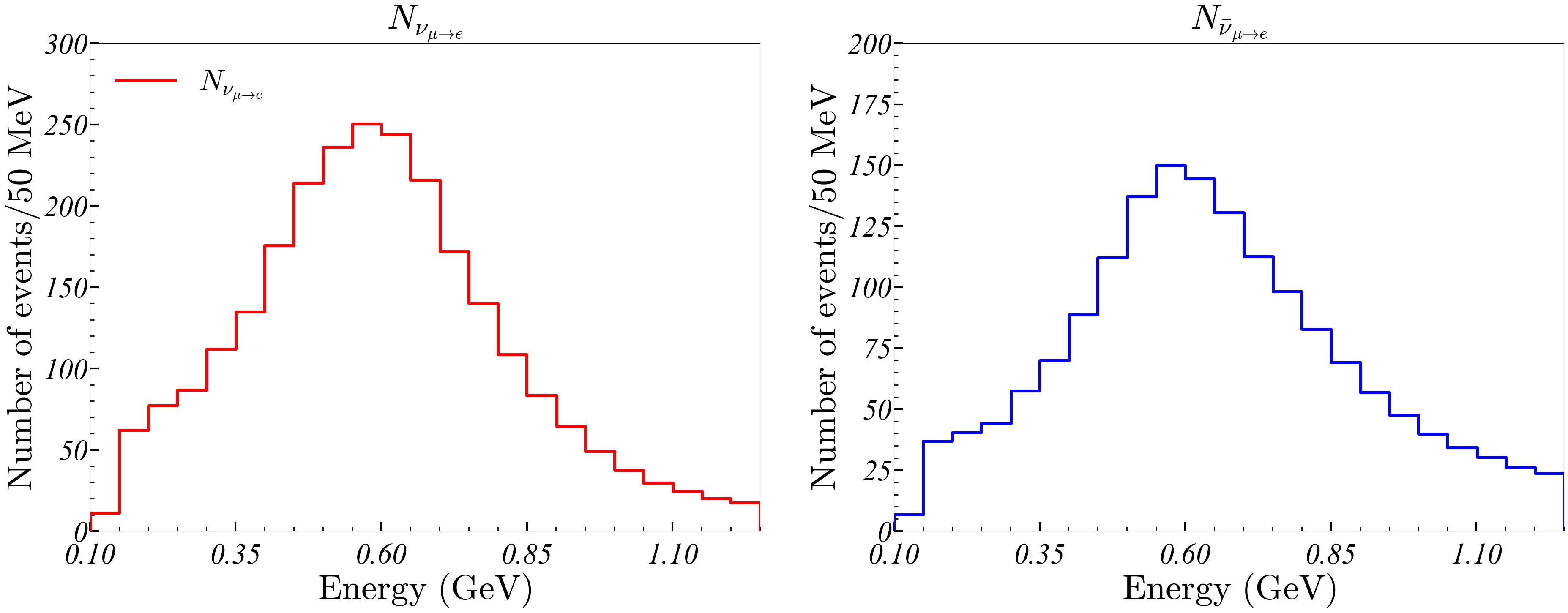}
    \caption{Number of events in the T2HK experiment. The left and right figures show the fluxes of $\nu_{\mu \to e}$ and $\bar{\nu}_{\mu \to e}$ at the far detector, respectively. Here, the CP phase is assumed to be $\delta_{\rm CP}=\ang{270}$. These expected number of events are calculated based on \cite{Hyper-Kamiokande:2018ofw}.}
    \label{fig:events-T2HK}
\end{figure}

In this study, we perform unitarity test assuming a situation where $\nu_\mu,~\bar{\nu}_\mu$, and $\nu_e$ beams are available.
The $\nu_\mu$ and $\bar{\nu}_\mu$ beams are available from the T2HK experiment~\cite{Hyper-Kamiokande:2018ofw}.
The off-axis beam is adopted to obtain neutrino beams that peak at 0.6 GeV.
The expected number of events of the $\nu_{\mu \to e}$ and $\bar{\nu}_{\mu \to e}$ channels are presented in Fig.~\ref{fig:events-T2HK}. Here we consider the charged current quasielastic interactions for the water Cherenkov detector and the neutrino oscillation.
The number of events is obtained based on Ref.~\cite{Hyper-Kamiokande:2018ofw}. We use the quasielastic neutrino cross sections from Ref.~\cite{Formaggio:2012cpf}.
The CP phase is assumed to be $\delta_{\rm CP}=\ang{270}$.

\begin{figure}[H]
    \centering
    \includegraphics[width=15cm]{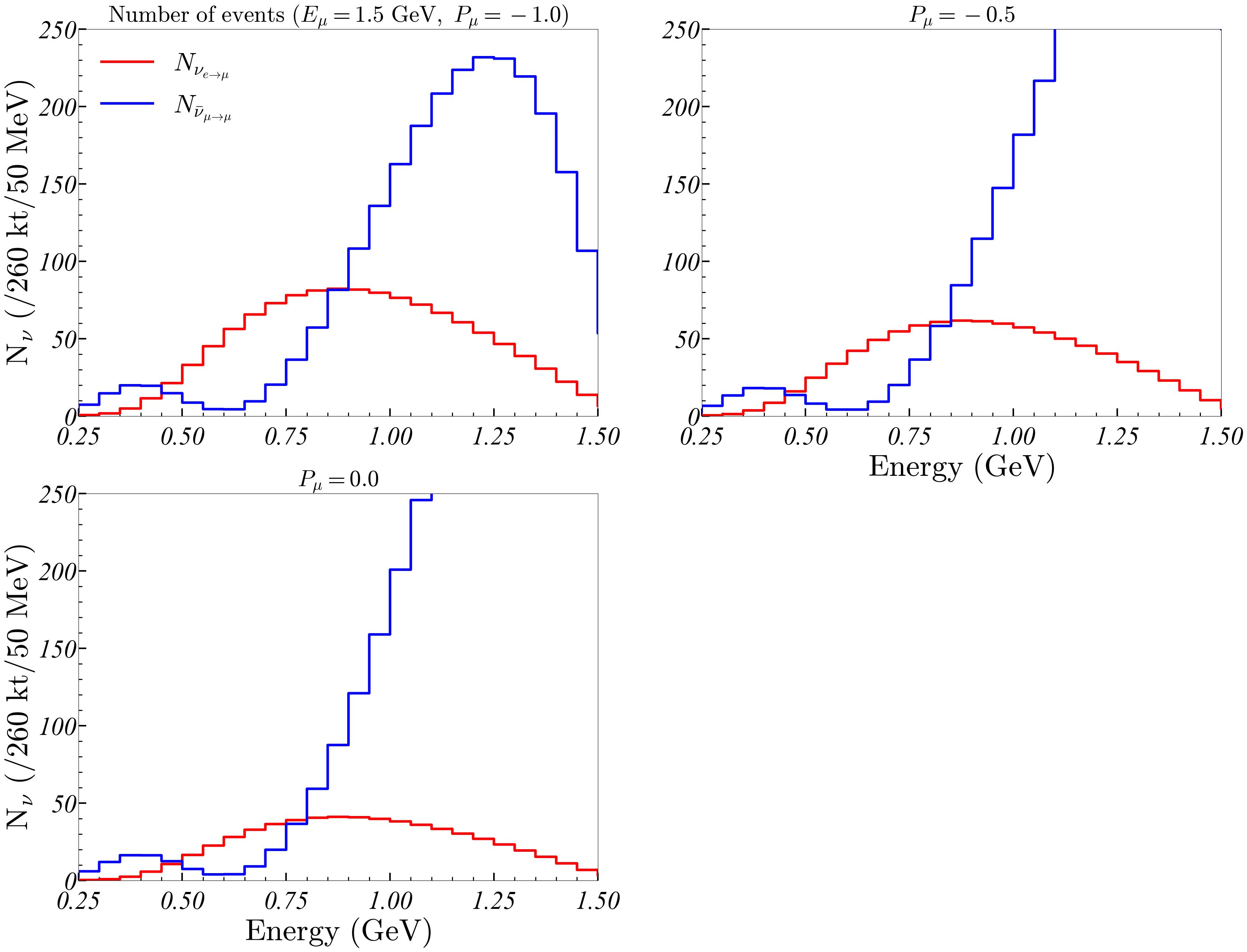}
    \caption{Neutrino flux measured at the Hyper-Kamiokande detector. The anti-muon beam energy is set to $1.5~\mathrm{GeV}$, and calculations are performed for three polarizations ($P_\mu=-1.0,\ -0.5,\ 0.0$). Here we set $\delta_{\rm CP}=\ang{270}$. The total number of muons is set to $10^{22}$.}
    \label{fig:events-NF}
\end{figure}

We can utilize the $\nu_e$ beam from the neutrino factory~\cite{Hamada:2022mua,Kitano:2024kdv,Kitano:2025wpc} at the J-PARC. The number of events can be estimated as follows~\cite{Geer:1997iz,Barger:1999fs}.
\begin{align}
    N_j^{\nu_{e \to \mu}} &= \int_{E_j}^{E_{j+1}} \frac{dE_\nu}{E_\mu} \times \frac{12N_\mu\cdot V_d\cdot n_N}{\pi L^2} \times \gamma^2 \qty(\frac{E_\nu}{E_\mu})^2 \times \qty[\qty(1-\frac{E_\nu}{E_\mu}) -P_\mu\qty(1-\frac{E_\nu}{E_\mu})] \notag \\
    & \times P^{\nu_{e\to\mu}}(E_\nu) \times \sigma_{\nu_\mu}(E_\nu)~,
    \label{eq:flux_e2m}
    \\
    N_j^{\bar{\nu}_{\mu \to \mu}} &= \int_{E_j}^{E_{j+1}} \frac{dE_\nu}{E_\mu} \times \frac{2N_\mu\cdot V_d\cdot n_N}{\pi L^2} \times \gamma^2 \qty(\frac{E_\nu}{E_\mu})^2 \times \qty[\qty(3-\frac{2E_\nu}{E_\mu}) -P_\mu\qty(1-\frac{2E_\nu}{E_\mu})] \notag \\
    & \times P^{\bar{\nu}_{\mu \to \mu}}(E_\nu) \times \sigma_{\bar{\nu}_\mu}(E_\nu)~,
    \label{eq:flux_mb2mb}
\end{align}
where $E_\mu$ is fixed muon energy, $N_\mu$ is total number of muons which decay towards the detector, $V_d$ is the detector volume, $n_N$ is the number density of the nucleon in water, and $L$ is the baseline length.
The $\gamma$ is the boost factor for muon. The polarization of the anti-muon, denoted by $P_\mu$, is an important factor affecting the $\nu_e$ flux.
In this analysis, we assume the combined analyses with the T2HK experiment. Thus, we consider that the baseline is $L=295~{\rm km}$ and the far detector is the Hyper-Kamiokande.   
We show the numbers of events, $N^{\nu_{e\to\mu}}$ and $N^{\bar{\nu}_{\mu\to\mu}}$, in Fig.~\ref{fig:events-NF}. In this case as well, we assume $\delta_{\rm CP}=\ang{270}$, and the events are smeared by the energy resolution ${\it \Delta}E=50~{\rm MeV}$~\cite{Hayato:2021heg}.
The values of anti-muon polarization are taken to be $P_\mu=-1.0,~-0.5,~0.0$. In addition, the muon energy is fixed to be $E_\mu=1.5~{\rm GeV}$, and the total number of muons is set to be $N_\mu=10^{22}$. 
%
In the Hyper-Kamiokande detector, when we observe neutrino beams from muon decay, the charges of the muons produced by charged-current interactions, $\nu_\mu n \to \mu^- p$ and $\bar{\nu}_\mu p \to \mu^+ n$, must be identified.
Namely, the $\bar{\nu}_\mu$ disappearance channel become a background of the $\nu_e$ appearance channel if the charges of muons cannot be identified at the Hyper-Kamiokande detector.
In fact, we can identify the charges by using the neutron tagging method~\cite{Beacom:2003nk,Huber:2008yx}, however, the efficiency is still under study~\cite{Super-Kamiokande:2023xup,Akutsu2019}.
Therefore, we consider three scenarios for the charge identification efficiency, $100\%$, $70\%$, and $0\%$. 
Due to this misidentification, we modify the appearance probability $P(\nu_{e\to\mu})$ and its statistical error as in Ref.~\cite{Kitano:2024kdv,Kitano:2025wpc}.  

\subsection{Unitarity test}
    In this analysis, we assume the experimental setup with $\nu_\mu,~\bar{\nu}_\mu$ from T2HK, and $\nu_e$ from a future neutrino factory at the J-PARC.
    We use energy range of $0.1\sim1.2~{\rm GeV}$ for T2HK and $0.25\sim1.45~{\rm GeV}$ for neutrino factory.
    Since the final states in the charged current interactions are different, we set a lower energy range for T2HK.

    We generate one million random virtual-experiments using a binomial distribution, and perform $\chi^2$ analysis to construct the new criterion $\tilde{\xi}$.
    We consider two types of virtual-experiments: one generated based on the standard three-generation model, and the other based on a four-generation model including an eV-scale sterile neutrino. 
    In this analysis, we adopt such a four-generation model as a simple example of the violation of $3\times3$ unitarity.
    We take the following reference values for the three-generation model, which are the arithmetic averages of the best fit points in Ref.~\cite{ParticleDataGroup:2022pth} from three groups, except for $\delta_{\rm CP}$. For the four-generation model, as a representative point, we refer to the results analyzed in Ref.~\cite{Dentler:2018sju,Parveen:2024bcc}, where all the twelve parameters are set to their best fit points.
    The chosen values of $\theta_{14}$ and $\theta_{24}$ are not excluded by the latest results from MicroBooNE~\cite{MicroBooNE:2025nll}.
    %
\begin{table}[H]
    \centering
    \begin{tabular}{c|c|c|c|c|c}
         $\theta_{12}$
         &$\theta_{13}$
         &$\theta_{23}$
         &$\delta_{\rm CP}$
         &$\mathit{\Delta} m_{21}^2/10^{-5}\ (\mathrm{eV}^2)$ 
         &$\mathit{\Delta} m_{31}^2/10^{-3}\ (\mathrm{eV}^2)$ 
        \\
         \hline 
         $\ang{33.9}$ & $\ang{8.49}$ & $\ang{48.1}$ & $\ang{270}$ & $7.43$ & $2.432$ \\
    \end{tabular}
    \caption{The reference values of three-generation oscillation parameters for the normal mass ordering.}
    \label{tab1}
\end{table}
\begin{table}[H]
    \centering
    \begin{tabular}{c|c|c|c|c|c}
         $\theta_{12}$
         &$\theta_{13}$
         &$\theta_{23}$
         &$\theta_{14}$
         &$\theta_{24}$ 
         &$\theta_{34}$ 
        \\
         \hline 
         $\ang{34.3}$ & $\ang{8.53}$ & $\ang{49.3}$ & $\ang{5.7}$ & $\ang{5}$ & $\ang{20}$ 
         \\
         \multicolumn{6}{c}{}
         \\
         \multicolumn{2}{c|}{$\delta_{\mathrm{CP}}$} & \multicolumn{2}{c|}{$\delta_{24}$} & \multicolumn{2}{c}{$\delta_{34}$}
         \\
         \hline
         \multicolumn{2}{c|}{$-\ang{165.6}$} & \multicolumn{2}{c|}{$\ang{0}$} & \multicolumn{2}{c}{$\ang{0}$}
         \\ 
         \multicolumn{6}{c}{}
         \\
         \multicolumn{2}{c|}{$\mathit{\Delta} m_{21}^2/10^{-5}\ (\mathrm{eV}^2)$} 
         & \multicolumn{2}{c|}{$\mathit{\Delta} m_{31}^2/10^{-3}\ (\mathrm{eV}^2)$} 
         & \multicolumn{2}{c}{$\mathit{\Delta} m_{41}^2\ (\mathrm{eV}^2)$}
         \\
         \hline
         \multicolumn{2}{c|}{$7.5$} & \multicolumn{2}{c|}{$2.55$} & \multicolumn{2}{c}{$1$}
    \end{tabular}
    \caption{The reference values of four-generation oscillation parameters for the normal mass ordering.}
    \label{tab2}
\end{table}

\begin{figure}[H]
    \centering
    \includegraphics[width=16cm]{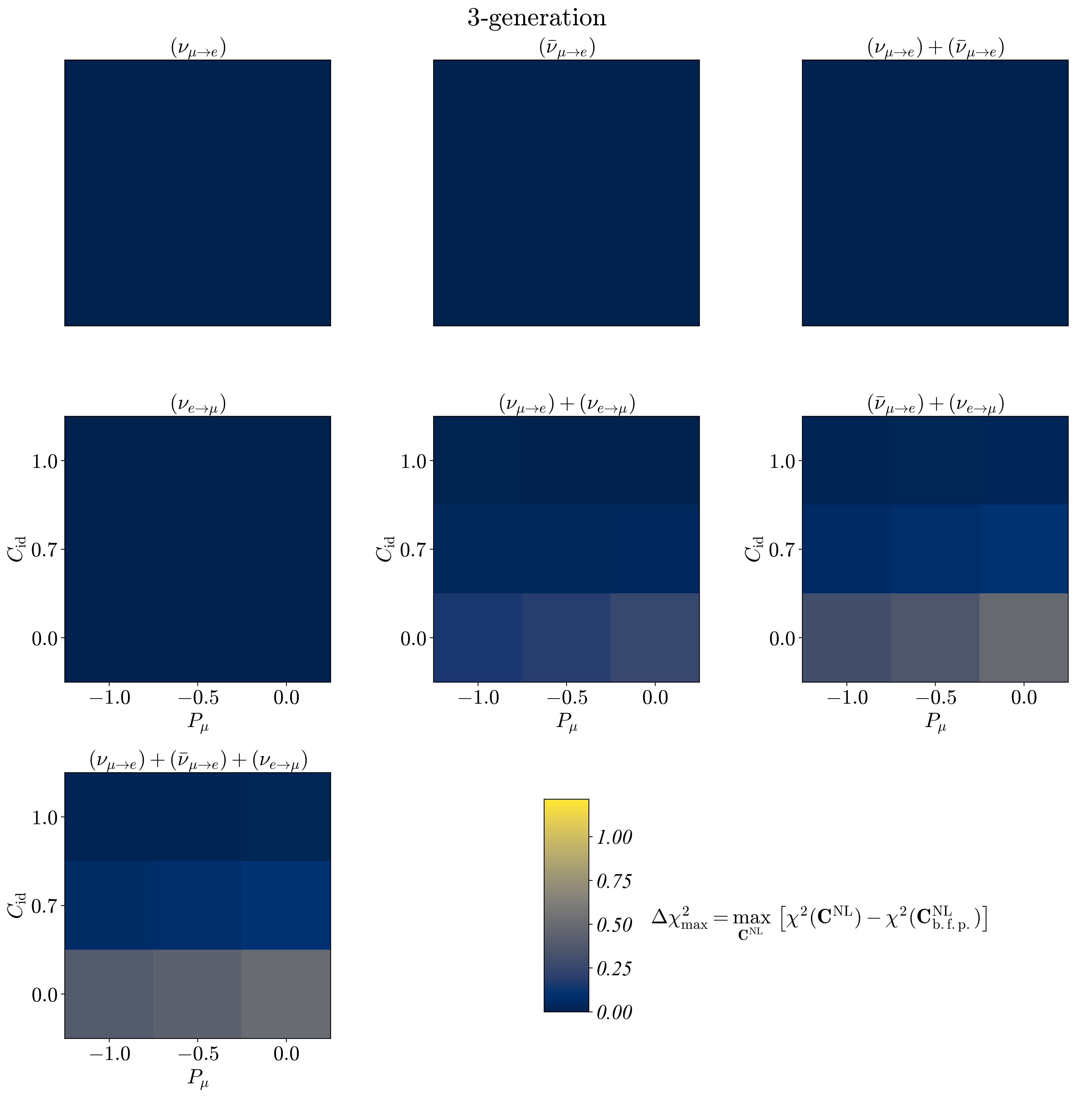}
    \caption{The values of ${\it \Delta}\chi^2_{\rm max}$. These results obtained in analyses where events generated by the three-generation model are fitted using the energy-dependent functions of three-generation model in matter. The non-linear parameters $C_5=X_1^{ee}$, $C_6=X_2^{ee}$, and $C_7=X_3^{ee}$ are varied within their $3\sigma$ ranges around the best fit points in Table~\ref{tab1}. Each figure presents results for both single-channel analyses and combinations of multiple channels. For analyses involving the $\qty(\nu_{e\to\mu})$ channel, cases with $C_{\rm id}=1.0,~0.7,~0.0$ and $P_\mu=-1.0,~-0.5,~0.0$ are also considered.
    The magnitude of ${\it \Delta}\chi^2_{\rm max}$ is shown using a color map.}
    \label{fig:dchi_3gen}
\end{figure}

\begin{figure}[H]
    \centering
    \includegraphics[width=16cm]{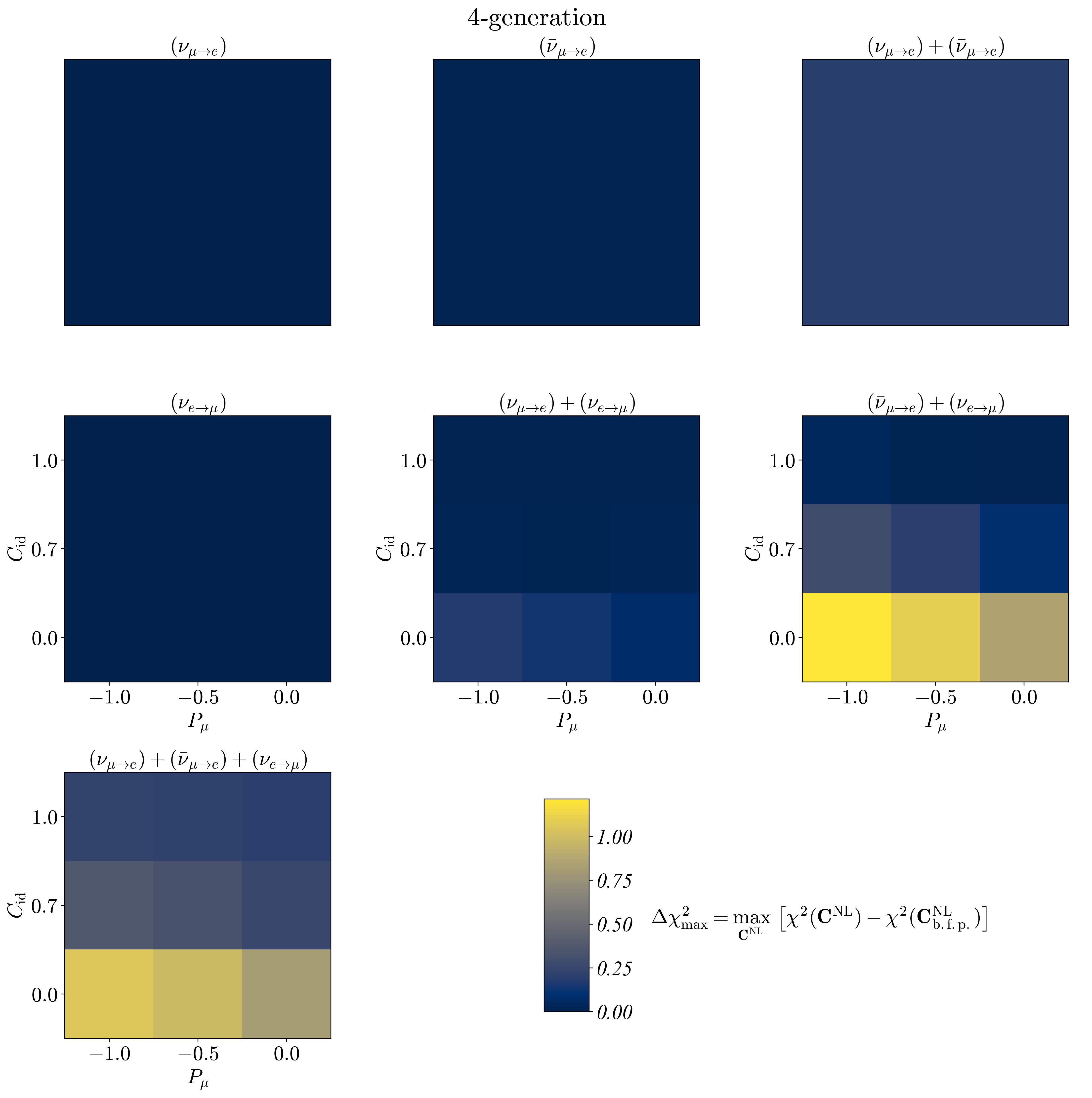}
    \caption{The values of ${\it \Delta}\chi^2_{\rm max}$. These results obtained in analyses where events generated by the three-generation model are fitted using the energy-dependent functions of four-generation model in matter. The non-linear parameters $C_5=X_1^{ee}$, $C_6=X_2^{ee}$, and $C_7=X_3^{ee}$ are varied within their $3\sigma$ ranges around the best fit points in Table~\ref{tab1}. Each figure presents results for both single-channel analyses and combinations of multiple channels. For analyses involving the $\qty(\nu_{e\to\mu})$ channel, cases with $C_{\rm id}=1.0,~0.7,~0.0$ and $P_\mu=-1.0,~-0.5,~0.0$ are also considered.
    The magnitude of ${\it \Delta}\chi^2_{\rm max}$ is shown using a color map.}
    \label{fig:dchi_4gen}
\end{figure}

Note that, in this analysis, we only construct the criterion $\tilde{\xi}$ by the extracted coefficients which are linear parameters.
As discussed in Sec.~\ref{sec:2}, given the non-linear parameters as inputs, a linear fit can be performed using Eq.~\eqref{eq:linear_fit}, while Eq.~\eqref{eq:chi2_non_linear} allows one to fit the non-linear parameters themselves. 
In this study, we consider T2HK and the neutrino factory, which have a baseline of $295~{\rm km}$.
At this level of baseline, neutrinos
would not go into deep in earth, and thus
matter effects
are not so significant yet non-negligible.
The sensitivities to the non-linear parameters
are, therefore, not quite large.
In other words, the value of $\chi^2$ varies only slightly around the best fit points of non-linear parameters in Table~\ref{tab1}.
In Figs.~\ref{fig:dchi_3gen} and \ref{fig:dchi_4gen}, we show the maximum ${\it \Delta}\chi^2 \equiv \chi^2\qty(\vb*{C}^{\rm NL}) - \chi^2\qty(\vb*{C}^{\rm NL}_{\rm b.f.p.})$ values when the non-linear parameters $C_5=X_1^{ee}$, $C_6=X_2^{ee}$, and $C_7=X_3^{ee}$ are varied within their $3\sigma$ ranges around the best fit points in Table~\ref{tab1}.
Figures~\ref{fig:dchi_3gen} and \ref{fig:dchi_4gen} show the maximum values of ${\it \Delta} \chi^2$, ${\it \Delta}\chi^2_{\rm max}$,
\begin{align}
    {\it \Delta}\chi^2_{\rm max} 
    = \max_{\vb*{C}^{\rm NL}}{\it \Delta}\chi^2
    = \max_{\vb*{C}^{\rm NL}} \qty[\chi^2\qty(\vb*{C}^{\rm NL}) - \chi^2\qty(\vb*{C}^{\rm NL}_{\rm b.f.p.})]~,
\end{align}
obtained in analyses where events generated by the three-generation and four-generation models are fitted using the energy-dependent functions of the three-generation model in matter.
In this analysis, since the fit is always performed using the three-generation model, the non-linear parameters are taken to be those of the three-generation case even when fitting events generated by the four-generation model.
Each figure presents results for both single channel analyses and combinations of multiple channels. For analyses involving the $\qty(\nu_{e\to\mu})$ channel, cases with $C_{\rm id}=1.0,~0.7,~0.0$ and $P_\mu=-1.0,~-0.5,~0.0$ are also considered.
The magnitude of ${\it \Delta}\chi^2_{\rm max}$ is shown using a color map.
As three parameters are varied, the statistical significance of ${\it \Delta}\chi^2_{\rm max}$ should be evaluated assuming a $\chi^2$ distribution with three degrees of freedom.
For instance, the value at the $68.3\%$ confidence level for a $\chi^2$ distribution with three degrees of freedom is about $3.53$; therefore, all ${\it \Delta}\chi^2_{\rm max}$ values in Figs.~\ref{fig:dchi_3gen} and \ref{fig:dchi_4gen} are statistically insignificant.
From these results, since varying the non-linear parameters around the best fit points does not yield a statistically significant difference, we fix the three non-linear parameters to their three-generation reference values and perform only a linear analysis in this study.
These results demonstrate that the sensitivity to unitarity violation, represented by $\tilde{\xi}$, is predominantly governed by the spectral shape of the oscillation probability rather than the precise values of the non-linear matter effect parameters. The fact that ${\it \Delta}\chi^2_{\rm max}$ remains statistically insignificant within the $3\sigma$ range confirms that any observed signal of $\tilde{\xi} \neq 0$ can be robustly interpreted as a consequence of non-unitarity in the mixing matrix. Therefore, fixing these parameters to their three-generation values is not merely a simplification, but a physically justified treatment that ensures our unitarity test remains independent of sub-leading matter effect uncertainties at this baseline.
This treatment greatly simplifies the analysis
for T2HK or similar baseline experiments
but a full non-linear fit would be necessary
if one consider the analysis in longer baseline
experiments.


In the following, the virtual-experiments generated by the three-generation model are referred to as ``three-generation events", and those generated by the four-generation model as ``four-generation events".
In this study, all analyses are performed by fitting with the three-generation model, Eq.~\eqref{eq:osc_prob_general}, even for the four-generation events. 
We show the results of testing unitarity in Sec.~\ref{sec:3_vac2mat} and \ref{sec:3_mat2mat}. First, in Sec.~\ref{sec:3_vac2mat}, we present the results of fitting virtual-experiments with matter effects using the energy dependence of the neutrino oscillation in vacuum. Then, in Sec.~\ref{sec:3_mat2mat}, we present the results of fitting such virtual-experiments using energy-dependent functions in matter.
We compare the best fit value of $\tilde{\xi}$ and its uncertainty in each case to evaluate the feasibility of the unitarity test.
The distribution of $\tilde{\xi}$ obtained from this analysis can be interpreted such that a larger deviation from $\tilde{\xi}=0$ indicates the violation of unitarity.
In the following figures, we compare the fit results for a single neutrino oscillation channel with those for combinations of multiple channels. 
We also show the results for different choices of the anti-muon beam polarization and the charge identification efficiency.
The polarization is set to $P_\mu=-1.0,~-0.5,~0.0$, and the charge identification efficiency is set to $C_{\rm id}=1.0,~0.7,~0.0$.
From the distribution of one million fitting results, the $68.3\%,~95.4\%$, and $99.7\%$ regions are represented by different color intensities.
We assume that the total number of muons (which decay towards the Hyper-Kamiokande detector) in the neutrino factory is $N_\mu=10^{22}$.

The distributions of the minimum of the $\chi^2$, denoted by $\chi^2_{\rm min}$, are presented in Appendix~\ref{sec:appendix:chi2}.  
Since the analysis using $10^6$ virtual-experiments provides the distribution of the $\chi^2_{\rm min}$ (i.e., the $\chi^2$ at the best fit point), the significance cannot be evaluated in the usual way as in standard $\chi^2$ analyses.
In this case, the statistical power can be evaluated from the distribution of $\chi^2_{\rm min}$. (See Appendix~\ref{sec:appendix:chi2} for a detailed explanation of the figure.)

\subsubsection{Fitting with Vacuum Formula to Events with Matter Effects}
\label{sec:3_vac2mat}
\begin{figure}[H]
    \centering
    \includegraphics[width=16cm]{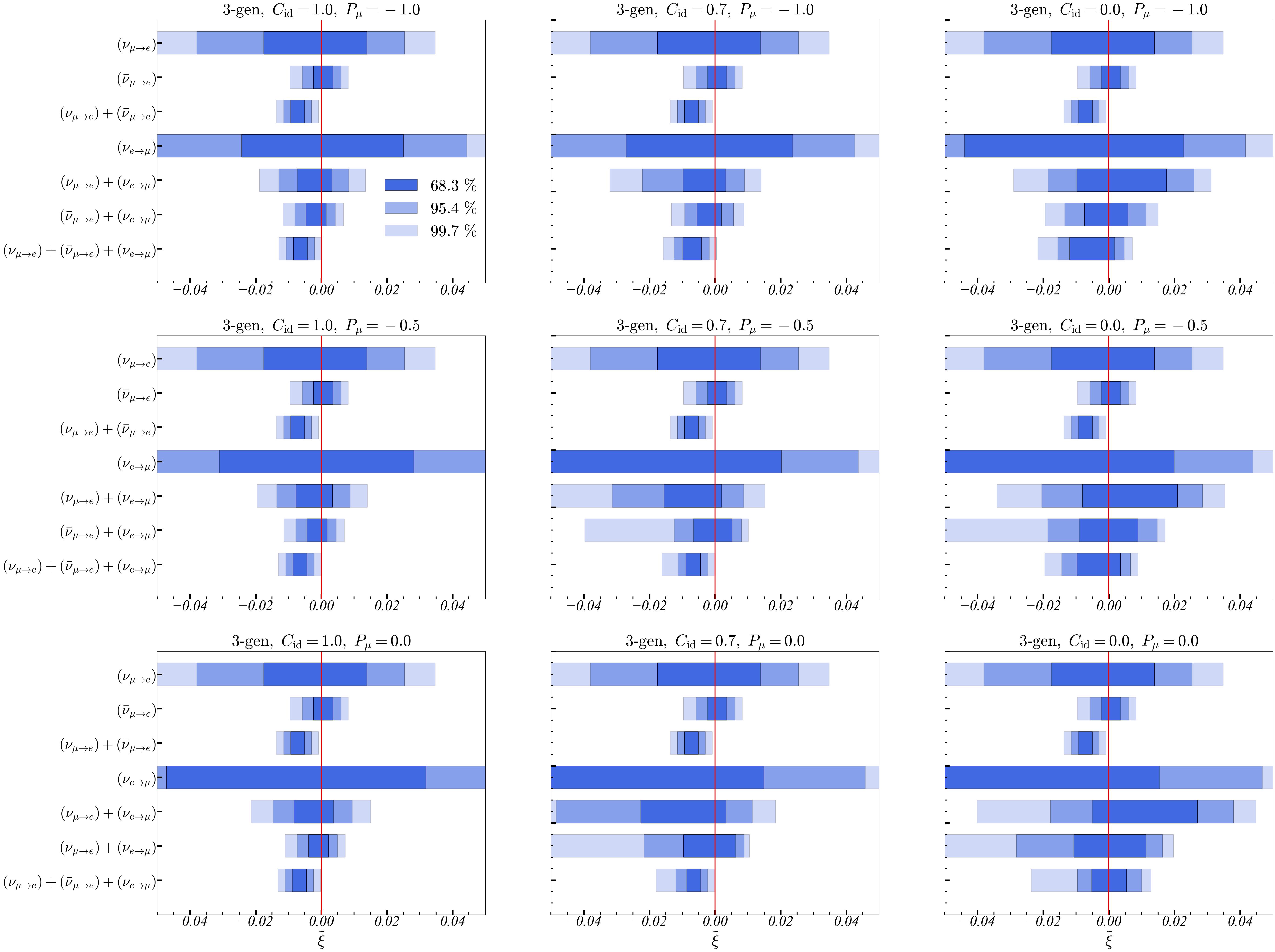}
    \caption{Unitarity test of the three-generation events
    in matter
    in terms of $\tilde{\xi}$ by ignoring matter effects
    in the fitting. These figures show some comparisons of various channel combinations. These figures also compare different values of anti-muon beam polarization~$P_\mu$ and charge identification efficiency~$C_{\mathrm{id}}$. From top to bottom, the figures show the cases with $P_\mu=-1.0,\ -0.5,\ 0.0$. From left to right, the figures show the cases with $C_{\mathrm{id}}=1.0,\ 0.7,\ 0.0$. In these figures, the total number of muons is set to $10^{22}$. The blue regions represent the results of the fitting three-generation model in vacuum to three-generation events in matter, with the color intensity indicating the allowed region up to the $3\sigma$ level. The red solid vertical lines represent $\tilde{\xi}=0$.}
    \label{fig:3gen-vac2mat}
\end{figure}

\begin{figure}[H]
    \centering
    \includegraphics[width=16cm]{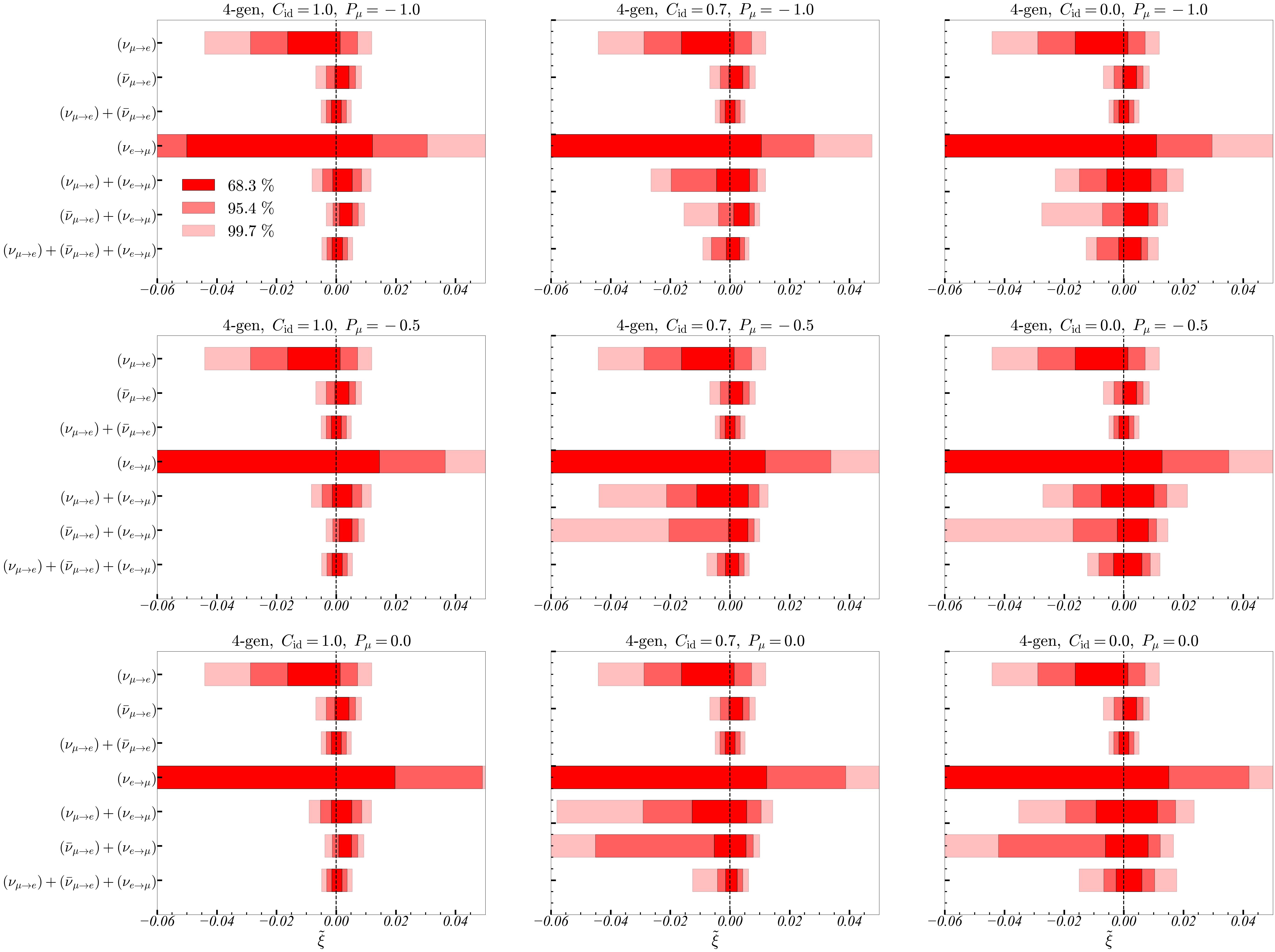}
    \caption{Unitarity test of the four-generation events
    in matter
    in terms of $\tilde{\xi}$ by ignoring matter effects
    in the fitting. As in Fig~\ref{fig:3gen-vac2mat}, these figures show some comparisons of various channel combinations and also compare different values of anti-muon beam polarization~$P_\mu$ and charge identification efficiency~$C_{\mathrm{id}}$. In these figures, the total number of muons is set to $10^{22}$. The red regions represent the results of the fitting three-generation model in vacuum to four-generation events in matter, with the color intensity indicating the allowed region up to the $3\sigma$ level. The black dashed vertical lines represent $\tilde{\xi}=0$.}
    \label{fig:4gen-vac2mat}
\end{figure}

As the first trial, we ignore the matter effects in the fitting functions, while the event generations are done 
with the matter effects. Of course, this is an inconsistent way of fitting, but 
we first observe the importance of including the matter effects in the analysis of the unitarity test.
In Fig.~\ref{fig:3gen-vac2mat} and \ref{fig:4gen-vac2mat}, we show the $\tilde{\xi}$ parameter 
obtained by fitting the three-generation model in vacuum to the virtual-experiment with matter effect. 
Figures~\ref{fig:3gen-vac2mat} and \ref{fig:4gen-vac2mat} show the results of fitting the  three-generation model in vacuum to the three-generation events in matter and to the four-generation events in matter, respectively.
The uncertainty varies depending on the channel or their combinations. In particular, when only ($\nu_{e\to\mu}$) channel from the neutrino factory is used, the uncertainty varies significantly depending on the polarization of the anti-muon beam and the charge identification efficiency.
However, combining it with channels from T2HK provides better results.
In the best case, $(C_{\rm id}=1.0,~P_\mu=-1.0)$, the channels, $(\nu_{\mu\to e})+(\bar{\nu}_{\mu\to e})$ from T2HK, and $(\nu_{\mu\to e})+(\bar{\nu}_{\mu\to e})+(\nu_{e\to\mu})$ combining T2HK and the neutrino factory, yield the best results.

In Fig.~\ref{fig:3gen-vac2mat}, for example, focusing on the analysis using $(\nu_{\mu\to e})$ and $(\bar{\nu}_{\mu\to e})$, the result shows that $\tilde{\xi} \neq 0$ at more than $3\sigma$.
In other words, even though the matter effects are small, neglecting them can lead to results that suggest 
a fake violation of unitarity.
On the other hand, in Fig.~\ref{fig:4gen-vac2mat}, although the fits are performed on four-generation events, the results appear to be consistent with unitarity.
Therefore, as in Fig.~\ref{fig:3gen-vac2mat}, one may obtain results that are inconsistent with the true choice of nature.
This is because the fits do not use energy-dependent functions that properly take matter effects into account.
However, for the T-conjugate combination $(\nu_{\mu\to e})+(\nu_{e\to\mu})$ (see also the distribution of $\chi^2_{\rm min}$ in the Appendix~\ref{sec:appendix:chi2}), 
no indication of the fake unitarity violation is observed. Thus, the T-conjugate combination can be regarded as insensitive to matter effects.
From the viewpoint of the $\chi^2_{\rm min}$ distribution, one cannot assign a single point relative to the expected distribution; instead, one can only state the probability (in percent) that the hypothesis can be excluded at a given significance level. As an example, in the $(\nu_{\mu\to e})+(\bar{\nu}_{\mu\to e})$ analysis shown in Fig.~\ref{fig:chi2-vac2mat}, the $\chi^2$ distribution expected under the correct model is shown by the black curve, and the corresponding significance levels are indicated by vertical lines. The colored histograms represent the distribution of $\chi^2_{\rm min}$ obtained with the setup of this section.
In this case, since the events in matter are fitted with the vacuum model, both distributions deviate from the expected distribution. Therefore, one can only state the probabilities of excluding the three-generation and four-generation model in vacuum at more than $3\sigma$, which are $46.64\%$ and $48.18\%$, respectively.
In the next section, we present the results of the analysis using the energy dependence of neutrino oscillation in matter.

%

\subsubsection{Fitting with Full Formula to Events with Matter Effects}
\label{sec:3_mat2mat}
\begin{figure}[H]
    \centering
    \includegraphics[width=16cm]{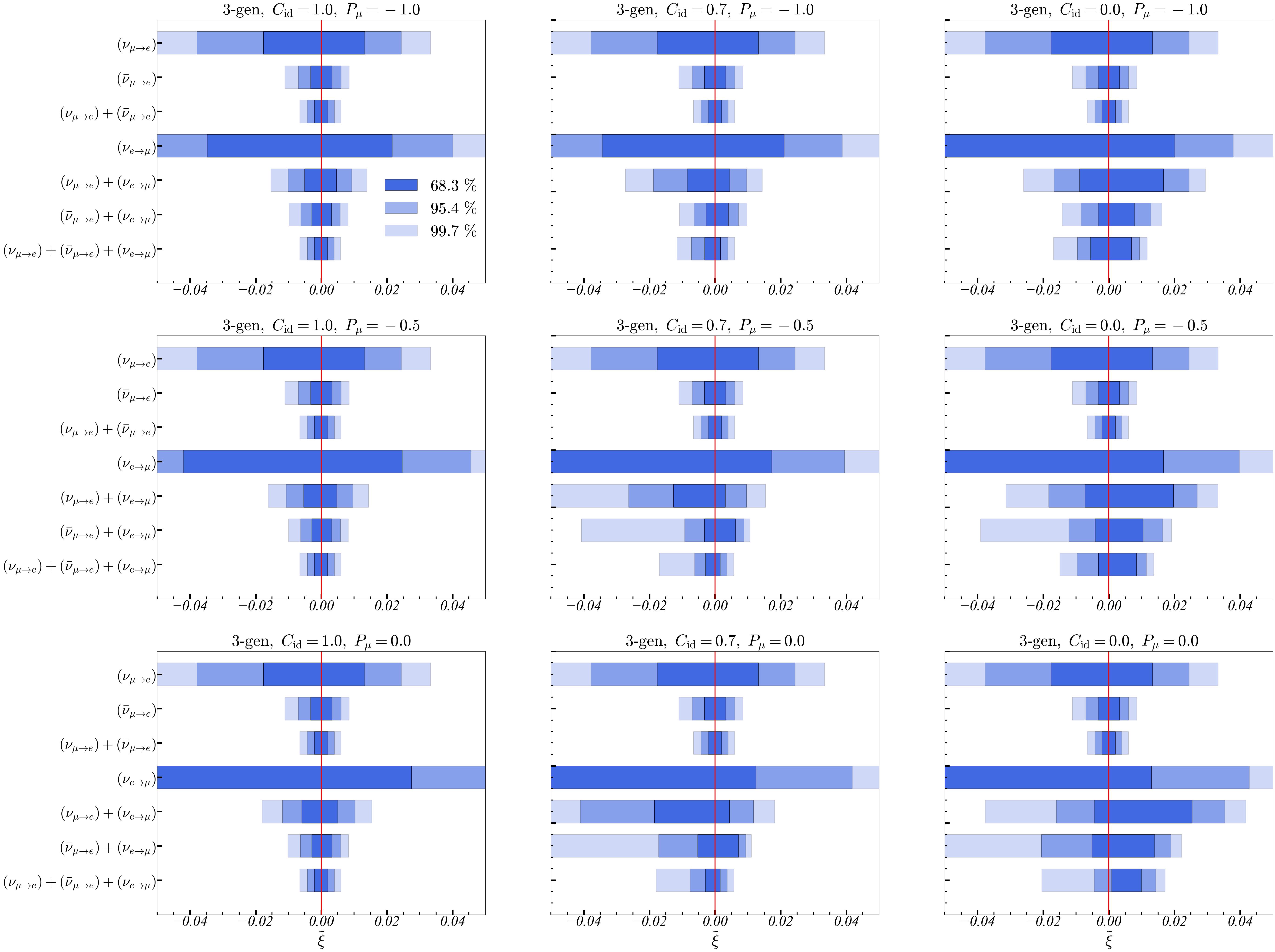}
    \caption{Unitarity test of the three-generation events
    in matter 
    in terms of $\tilde{\xi}$ by including matter effects in the fitting. These figures show some comparisons of various channel combinations. These figures also compare different values of anti-muon beam polarization~$P_\mu$ and charge identification efficiency~$C_{\mathrm{id}}$. From top to bottom, the figures show the cases with $P_\mu=-1.0,\ -0.5,\ 0.0$. From left to right, the figures show the cases with $C_{\mathrm{id}}=1.0,\ 0.7,\ 0.0$. In these figures, the total number of muons is set to $10^{22}$. The blue regions represent the results of the fitting three-generation model in matter to three-generation events in matter, with the color intensity indicating the allowed region up to the $3\sigma$ level. The red solid vertical lines represent $\tilde{\xi}=0$.}
    \label{fig:3gen-mat2mat}
\end{figure}

\begin{figure}[H]
    \centering
    \includegraphics[width=16cm]{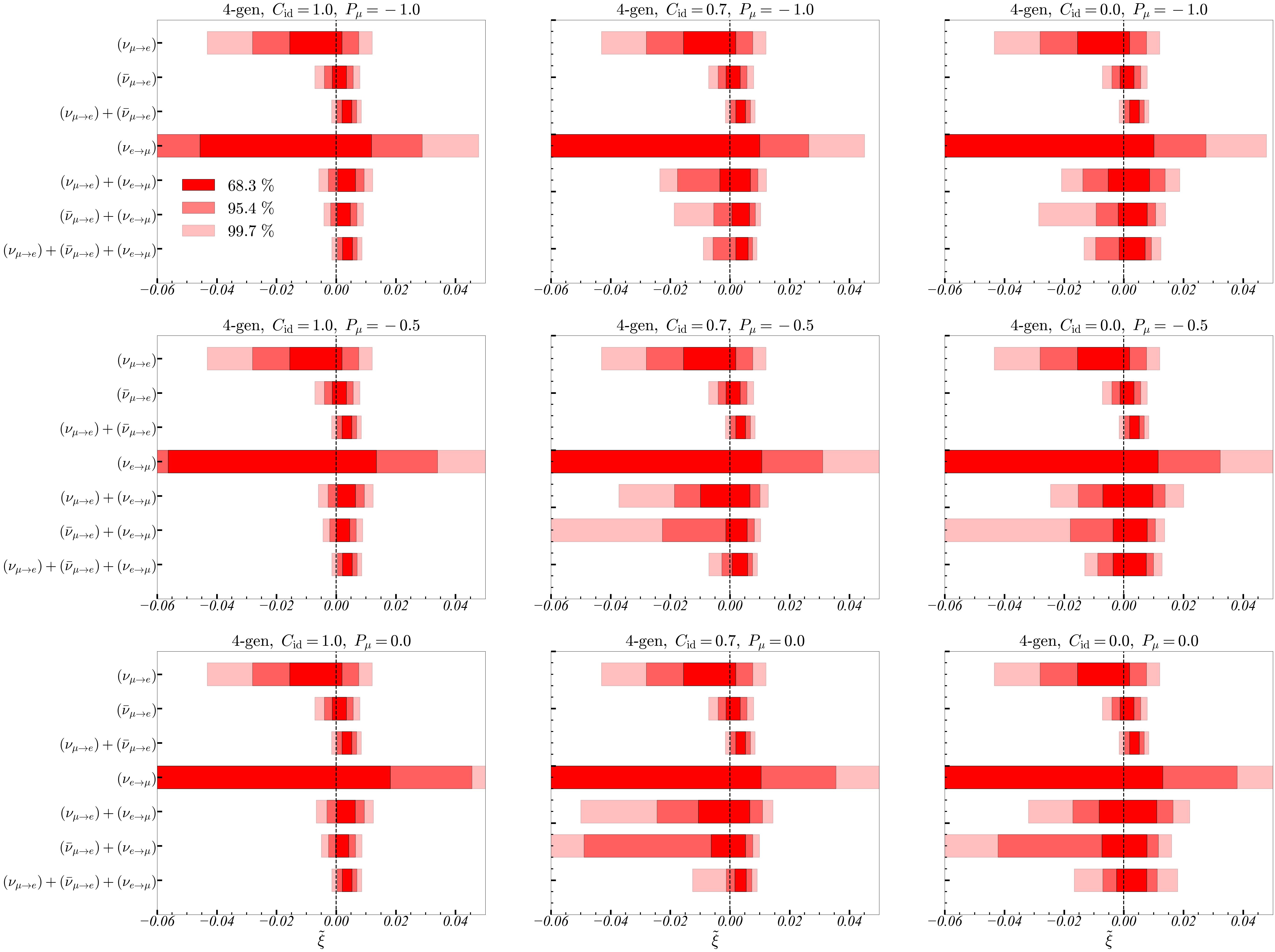}
    \caption{Unitarity test of the four-generation events
    in matter 
    in terms of $\tilde{\xi}$ by including matter effects in the fitting. As in Fig~\ref{fig:3gen-vac2mat}, these figures show some comparisons of various channel combinations and also compare different values of anti-muon beam polarization~$P_\mu$ and charge identification efficiency~$C_{\mathrm{id}}$. In these figures, the total number of muons is set to $10^{22}$. The red regions represent the results of the fitting three-generation model in matter to four-generation events in matter, with the color intensity indicating the allowed region up to the $3\sigma$ level. The black dashed vertical lines represent $\tilde{\xi}=0$.}
    \label{fig:4gen-mat2mat}
\end{figure}

Given all the tests, we now perform the real analysis by using 
the full fitting with matter effects to the events generated with
matter effects.
Figures~\ref{fig:3gen-mat2mat} and \ref{fig:4gen-mat2mat} show the results of fitting the  three-generation model in matter to the three-generation events in matter and to the four-generation events in matter, respectively.
In Fig.~\ref{fig:3gen-mat2mat}, almost all analyses show results consistent with $\tilde{\xi}=0$.
Compared with Fig.~\ref{fig:3gen-vac2mat}, one can confirm the importance
of the matter effects in fitting even in the case where the baseline is relatively short. 
In the worst case $(C_{\rm id}=0.0,~P_\mu=0.0)$, the analysis using $(\nu_{\mu\to e})+(\bar{\nu}_{\mu \to e})+(\nu_{e\to\mu})$, the best fit
deviates from $\tilde \xi = 0$ although it is not statistically significant.
For the test of unitarity,
the charge identification efficiency and the muon beam polarization
significantly help to reduce the uncertainties.

For the analysis of the four-generation events, 
one can see in Fig.~\ref{fig:4gen-mat2mat}
that the $\tilde \xi = 0$ can be excluded, i.e., the unitarity violation
is confirmed, at more than $2\sigma$ level
by using the CP conjugate channel $(\nu_{\mu\to e})+(\bar{\nu}_{\mu \to e})$ and 
also the one adding $(\nu_{e\to\mu})$ mode.
Even if we use the best fit values for the four-generation model in Ref.~\cite{Dentler:2018sju,Parveen:2024bcc}, the $2\sigma$ level exclusion
of unitarity is obtained by the $\tilde \xi$ analysis.
One can see that the charge identification efficiency $C_{\rm id}$ and the beam polarization $P_\mu$ are important particularly for the channel including the $(\nu_{e\to\mu})$ mode. In some cases, the $\tilde \xi = 0$ is 
consistent at $1\sigma$ level even though the underlying theory is not unitary.

Instead of $\tilde{\xi}$, when considering the distribution of $\chi^2_{\rm min}$ (see Fig.~\ref{fig:chi2-mat2mat} in Appendix~\ref{sec:appendix:chi2}), the fit to three-generation events in matter, using the energy dependence of the three-generation model in matter, is in good agreement with the expected distribution, whereas the corresponding fit to four-generation events shows a deviation.
However, no difference at the level of more than $2\sigma$, as seen in $\tilde{\xi}$, is observed. For example, in the analysis using $(\nu_{\mu\to e})+(\bar{\nu}_{\mu\to e})$, one can only state that the probability of excluding the four-generation model at more than $2\sigma$ is $23.24\%$.
In this case, the $\tilde \xi$ analysis can provide a much more powerful test of unitarity than the fit-based approach.

%
%

\section{Summary}
\label{sec:4}
In this study, we extend the method for testing unitarity, originally proposed for vacuum neutrino oscillations in Ref.~\cite{Kitano:2025wpc}, to the case of matter.
We then apply the method and evaluate the sensitivity in future experiments.
In this analysis, we assume T2HK~\cite{Hyper-Kamiokande:2018ofw} and the neutrino factory with a $\nu_e$ beam at J-PARC~\cite{Hamada:2022mua,Kitano:2024kdv,Kitano:2025wpc}, and test unitarity using appearance channels of neutrino oscillation.
When we consider the matter effects in neutrino oscillations, the oscillation probability depends on both linear and non-linear parameters.
As shown in Figs.~\ref{fig:dchi_3gen} and \ref{fig:dchi_4gen}, even when the three non-linear parameters are varied within their $3\sigma$ ranges around the best fit points, ${\it \Delta}\chi^2_{\rm max}$ does not show a statistically significant change. Therefore, we fix the three non-linear parameters to their best fit values in the three-generation case and perform a linear analysis.

First, we fit the three-generation model in vacuum to three- or four-generation events in matter, and extract the four parameters in Eq.~\eqref{eq:osc_prob_general}.
In Sec.~\ref{sec:3_vac2mat}, from the fit of the three-generation model in vacuum to three-generation events in matter, it is found that the unitarity appears to be violated when using $(\nu_{\mu\to e})+(\bar{\nu}_{\mu\to e})$ or $(\nu_{\mu\to e})+(\bar{\nu}_{\mu\to e})+(\nu_{e\to\mu})$.
In contrast, the fit of the three-generation model in vacuum to four-generation events in matter suggests that the unitarity is not violated.
Therefore, we find that the matter effects must be considered even in the case of relatively short baseline.

In Sec.~\ref{sec:3_mat2mat}, we present the analysis using the energy dependence of neutrino oscillations in matter.
Unlike the case of the three-generation vacuum model, when the energy-dependent functions in matter are taken into account, the results show that $\tilde{\xi}=0$. 
When we consider that the four-generation parameters are best fit points, we find that analyses based on channel $(\nu_{\mu\to e})+(\bar{\nu}_{\mu\to e})$, as well as those including channel $(\nu_{\mu\to e})+(\bar{\nu}_{\mu\to e})+(\nu_{e\to\mu})$, can provide a sensitivity more than $2\sigma$ to the unitarity violation.

Our analysis elucidates that the synergy between the CP-conjugate channels at T2HK and the T-conjugate channels at future neutrino factories is not merely a matter of increasing statistics, but a fundamental requirement for a robust unitarity test. 
While the CP-conjugate combination ($\nu_{\mu\to e})+(\bar{\nu}_{\mu\to e}$) is highly sensitive to the leptonic CP phase, it is intrinsically sensitive to asymmetries induced by matter effects, which can mimic unitarity violation. 
In contrast, the T-conjugate combination ($\nu_{\mu\to e})+(\nu_{e\to\mu}$) is found to be robust against uncertainties in matter effects. 
By combining these complementary channels, we can effectively separate the signals of non-unitarity from the matter effects, providing a `clean' probe of the PMNS matrix even with relatively short baselines.

Reference~\cite{Kitano:2025wpc} considered only the vacuum case and briefly discussed, in the summary, the impact of matter effects on T- and CP-conjugate channels. 
In this paper, by fitting events in matter with the energy dependence of neutrino oscillations in both vacuum and matter, we show that matter effects cannot be neglected.
Our results emphasize that even for relatively short baselines, such as 295 km, the systematic shift induced by matter effects is large enough to be misidentified as a $3\sigma$ signal of unitarity violation if vacuum formulas are used. 
This potential for the `false positive' highlights that the experimental sensitivity to new physics is strictly limited by our treatment of matter effects. 
Therefore, the implementation of our analysis method including matter effects is not merely an improvement in precision, but a necessary requirement to ensure that any claimed discovery of non-unitarity truly comes from the lepton mixing structure.
We also find that analysis using T-conjugate channel is insensitive to matter effects.
However, since only statistical uncertainties are considered in this analysis, systematic uncertainties must also be included to make the analysis more feasible.

Lepton mixing is a key to probing the fundamental nature of neutrinos.
The unitarity test of lepton mixing is an important task in future neutrino oscillation phenomenology.
In this study, we evaluate the possibility of testing unitarity in future long baseline experiments and show that independent measurements at T2HK and future neutrino factories are well motivated.  
%

\section*{Acknowledgements}
This work is supported in part
by JST SPRING Japan Grant Number JPMJSP2178 (SS).
This work is also supported in part by JSPS KAKENHI Grant-in-Aid for
Scientific Research (No.~22K21350~[RK], No.~25H01524~[JS], No.~26K21729~[JS]) and
the U.S.-Japan Science and Technology Cooperation Program in High Energy Physics~(2025-20-2~[RK]).

\appendix
\section{Distribution of $\chi^2_{\rm min}$}
\label{sec:appendix:chi2}
\begin{figure}[H]
    \centering
    \includegraphics[width=16cm]{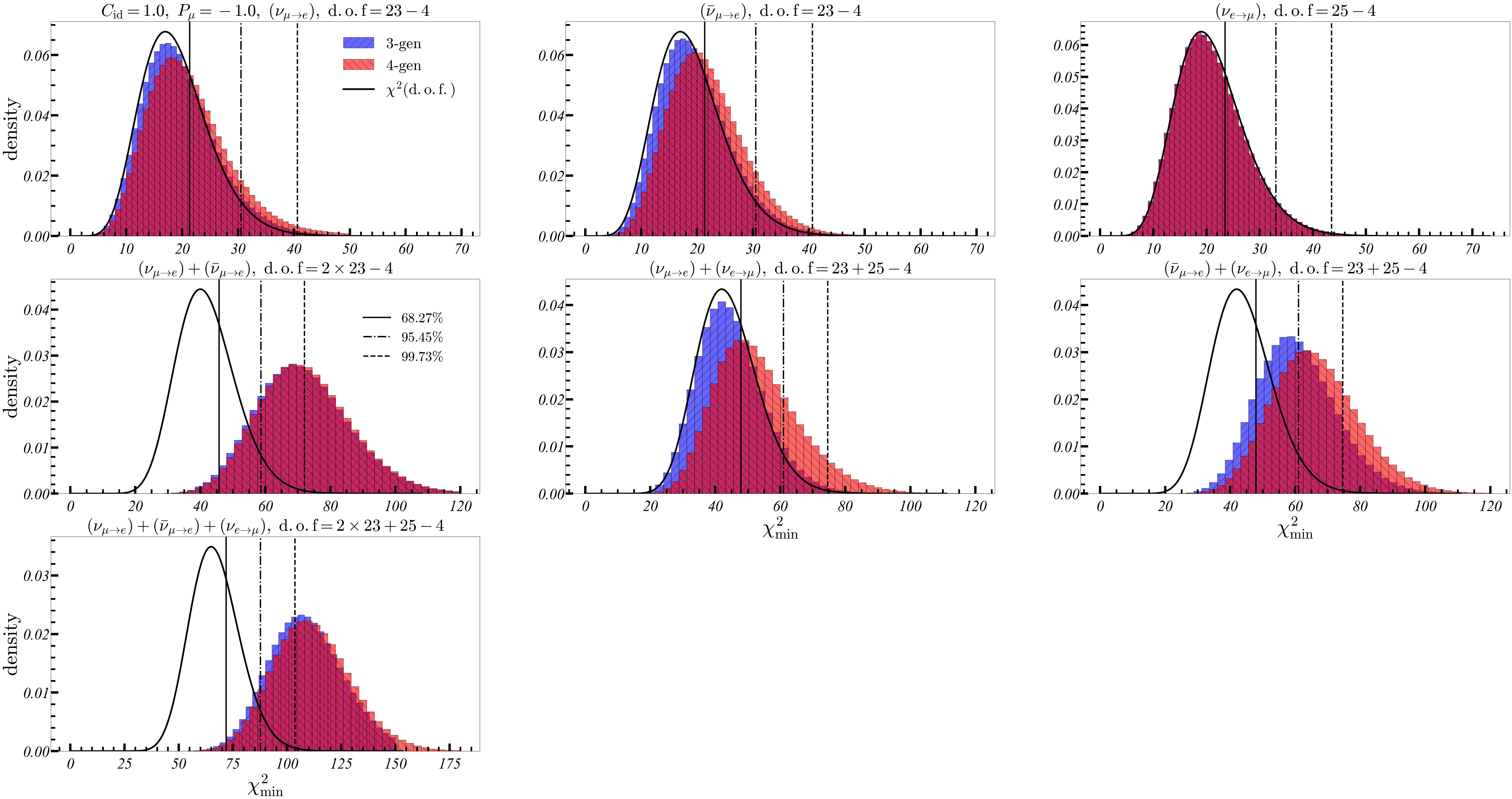}
    \caption{Distribution of $\chi^2_{\rm min}$. These figures show some comparisons of various channel combinations. The figures show the case with $C_{\rm id}=1.0,~P_\mu=-1.0$, and $N_\mu=10^{22}$. The black curves represent the probability density functions of the $\chi^2$ distribution for each degree of freedom (see also the blue histogram in Fig.~\ref{fig:chi2-mat2mat} for comparison). The blue histogram shows the $\chi^2_{\rm min}$ distribution obtained by fitting the three-generation model in vacuum to the three-generation events in matter, while the red histogram shows the $\chi^2_{\rm min}$ distribution obtained by fitting the three-generation model in vacuum to the four-generation events in matter. The vertical lines (solid, dash-dotted, and dashed) correspond to significance level of $68.27\%,~95.45\%$, and $99.73\%$, respectively, in the $\chi^2$ distribution shown by the black curves.}
    \label{fig:chi2-vac2mat}
\end{figure}

\begin{figure}[H]
    \centering
    \includegraphics[width=16cm]{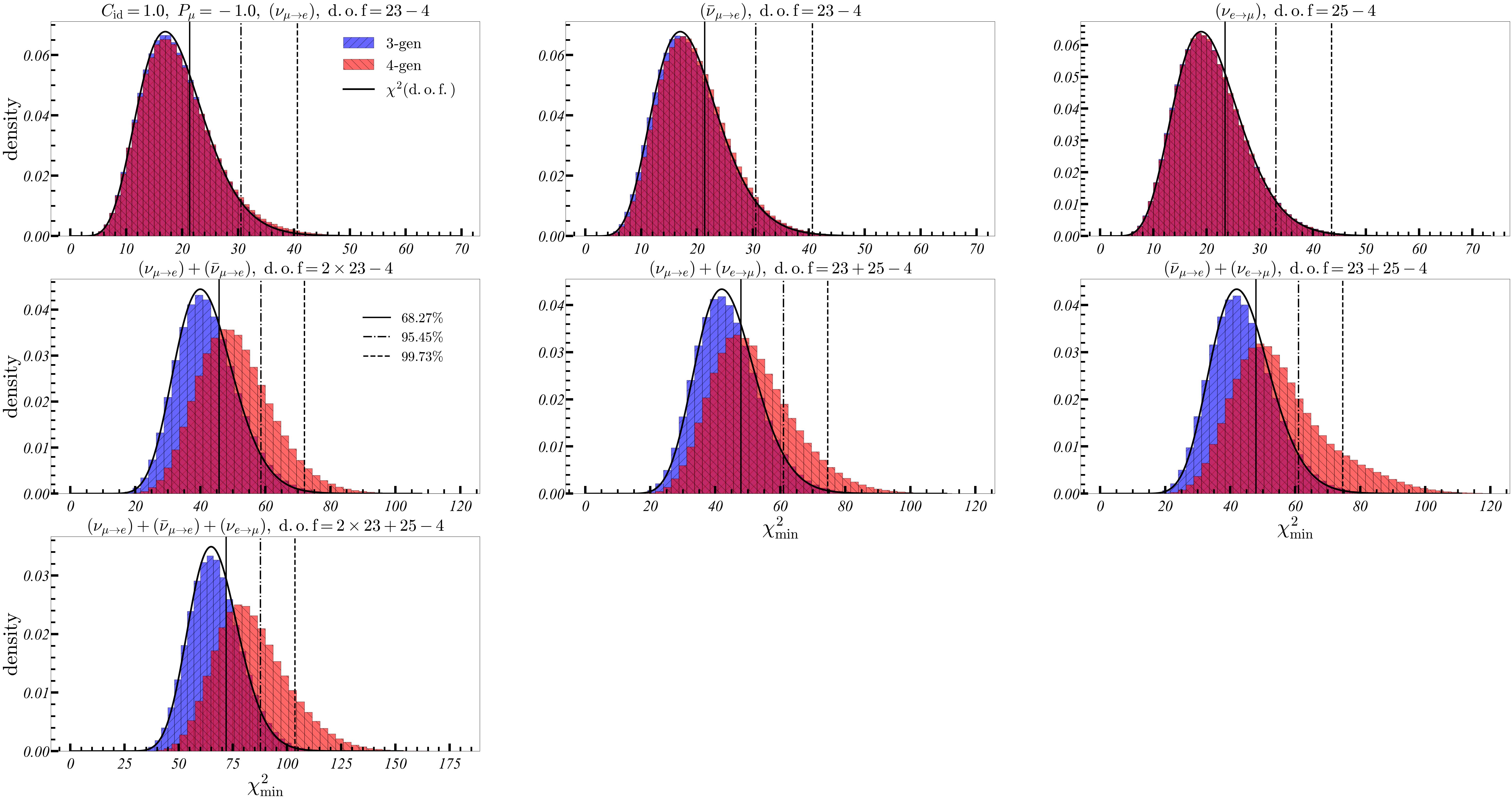}
    \caption{Distribution of $\chi^2_{\rm min}$. These figures show some comparisons of various channel combinations. The figures show the case with $C_{\rm id}=1.0,~P_\mu=-1.0$, and $N_\mu=10^{22}$. The black curves represent the probability density functions of the $\chi^2$ distribution for each degree of freedom. The blue histogram shows the $\chi^2_{\rm min}$ distribution obtained by fitting the three-generation model in matter to the three-generation events in matter, while the red histogram shows the $\chi^2_{\rm min}$ distribution obtained by fitting the three-generation model in matter to the four-generation events in matter. The vertical lines (solid, dash-dotted, and dashed) correspond to significance level of $68.27\%,~95.45\%$, and $99.73\%$, respectively, in the $\chi^2$ distribution shown by the black curves.}
    \label{fig:chi2-mat2mat}
\end{figure}

In this section, we present the distribution of $\chi^2_{\rm min}$ analyzed by Eq.~\eqref{eq:def_chi2}.
Figures~\ref{fig:chi2-vac2mat} and \ref{fig:chi2-mat2mat} show the distribution of $\chi^2_{\rm min}$ for each channel obtained from one million virtual-experiments in the case of $C_{\rm id}=1.0$ and $P_\mu=-1.0$.
The black solid curves represent the $\chi^2$ distributions for the corresponding degrees of freedom, while the blue (red) histograms show the results of fits of the three-generation model to the three-generation (four-generation) events.
The vertical solid, dash-dotted, and dashed lines correspond to the lower $68.27\%,~95.45\%$, and $99.73\%$ points of the $\chi^2$ distribution represented by the black solid curves, respectively.
Figure~\ref{fig:chi2-vac2mat} shows the results of fits using the energy dependence in vacuum.
On the other hand, the results of Fig.~\ref{fig:chi2-mat2mat} correspond to the fit by the energy dependence in matter.  
Figures~\ref{fig:chi2-vac2mat} and \ref{fig:chi2-mat2mat} compare the distribution under the null and alternative hypotheses in the context of statistical hypothesis testing, allowing one to evaluate the statistical power of the test.
The $\chi^2$ distributions represented by the black curves in Figs.~\ref{fig:chi2-vac2mat} and \ref{fig:chi2-mat2mat} correspond to the null hypothesis, whereas the histograms correspond to the alternative hypothesis.

In general statistical tests in experiments, the significance is determined by where the observed statistical quantity lies within the null distribution.
However, in practice, the alternative hypothesis itself has a distribution, and therefore a Type II error exists. 
Since we cannot know the true distribution, the probability of the Type II error, i.e., the probability that the histogram lies to the left of the vertical line indicating the significance level.
Conversely, by evaluating the probability of the alternative hypothesis lying to the right of the significance threshold, one can estimate the probability of excluding it at a given significance level.

Since Fig.~\ref{fig:chi2-vac2mat} shows the results of fitting the three-generation model in vacuum to the three- or four-generation events in matter, the histograms must be excluded. 
In other words, it is desirable for the distribution of the alternative hypothesis to be well separated from that of the null hypothesis, as this leads to a smaller probability of a Type II error (i.e., a larger statistical power).

In these results as well, combining multiple channels leads to statistically more favorable outcomes.
Focusing on the blue histogram for the analysis using $(\nu_{\mu\to e})+(\nu_{e\to\mu})$, the null and alternative distributions almost overlap, despite the fact that the fit is incorrect.
This channel corresponds to a T-conjugate channel, and the results indicate that such channels are insensitive to matter effects.

Since Fig.~\ref{fig:chi2-mat2mat} shows the results of fitting the three-generation model in matter to the three- or four-generation events in matter, the blue histograms are expected to overlap with the null distribution, while the red histograms are expected to be well separated from it.
In this figure, unlike Fig.~\ref{fig:chi2-vac2mat}, the blue distribution corresponds to the result of correct fits and indeed overlaps with the null distribution.
On the other hand, in the analyses combining multiple channels, which are statistically more favorable, the red distribution is well separated from the null distribution.

Finally, in the analysis based on $\tilde{\xi}$ as in Figs.~\ref{fig:3gen-vac2mat}, \ref{fig:4gen-vac2mat}, \ref{fig:3gen-mat2mat}, and \ref{fig:4gen-mat2mat}, a distribution of $\tilde{\xi}$ corresponds to an alternative hypothesis, while $\tilde{\xi}=0$ corresponds a null hypothesis. 
Therefore, we can evaluate the significance by determining where $\tilde{\xi}=0$ lies within the $\tilde{\xi}$ distribution.
While the extraction of $\tilde{\xi}$ relies on the $\chi^2$ fitting procedure, the resulting analysis provides a more direct probe of unitarity than a test based solely on the $\chi^2_{\rm min}$ distribution.


\bibliographystyle{./utphys.bst}
\bibliography{./UnitarityTestMatter.bib}

\end{document}